\documentclass[doublecol]{epl2}

\usepackage{lineno,hyperref}
\usepackage{lineno}
\usepackage{graphicx}
\usepackage{subcaption}
\usepackage{amsmath}
\usepackage{verbatim}
\usepackage{amssymb}
\usepackage{color}
\usepackage{mathrsfs}

\title{Kink scattering in a Lorentz-violating $\phi^6$ model}
\shorttitle{Kink scattering in a Lorentz-violating $\phi^6$ model} 

\author{Haobo Yan\footnote{E-mail: haobo@stu.pku.edu.cn (corresponding author)}}
\shortauthor{Haobo Yan}

\institute{
School of Physics, Peking University, Beijing 100871, China \\
School of Physics, Xi’an Jiaotong University, Xi’an 710049, China
}

\abstract{The role played by a Lorentz-violating term on the outcomes of kink scattering in the $\phi^6$ model is investigated by using the Fourier spectral method. Impacts of the Lorentz-violating term on the critical velocities, the location of two-bounce windows, and the maximal values of various types of energy densities are analyzed. Some novel features of kink-antikink collisions are discussed. The interactions between three and four kinks are also considered.}

\begin{document}

\maketitle


\section{Introduction}
In $(1+1)$-dimensional nonlinear scalar field models with more than one degenerated vacua, there may exist a type of classical solution called kink, which smoothly connects different vacua and has localized energy density. In the past decades, kink and its higher-dimensional extension, domain wall, have been extensively studied in both condensed matter physics and high energy physics~\cite{Vachaspati2006}.

A related topic that has been repeatedly discussed recently, is the scattering of non-integrable kinks. Unlike the collision of integrable kinks where the kinks simply pass through each other, a pair of incoming non-integrable kinks (for instance, $\phi^4$ kinks) can be bounced back after a finite time of the collision or form a bound state called bion/oscillon, depending on their initial velocities, say $v_0$ and $-v_0$ for symmetric collisions~\cite{Kudryavtsev1975, Sugiyama1979, Moshir1981, Campbell1983}. The velocity intervals within which kinks are bounced back after $n$ collisions are called $n$-bounce windows ($n$BWs). For the $\phi^4$ model, the $n$-bounce windows form an interesting fractal structure~\cite{Anninos1991}.

Bounce windows have also been found in many other non-integrable kink models~\cite{Antunes2004,Hoseinmardy2010,Dorey2011,Gomes2014,Khare2014,Arthur2016,Lima2019,Belendryasova2019,Christov2019,Khare2019,Adam2020,Mohammadi2021,AlonsoIzquierdo2021,Nzoupe2021,Dorey2021,Campos2021,Bazeia2020,Zhong2020,Christov2021}. A well-accepted explanation for the appearance of bounce windows is the resonant energy exchange between the translational and vibrational modes of the kink~\cite{Campbell1983,Dorey2011}. This idea is shown recently using a collective coordinate model~\cite{Manton2021b}. In addition to double kink collisions, the simultaneous collisions of multi-kink at the same point have also been investigated in many works~\cite{Gani2019, MoradiMarjaneh2017, Ekomasov2018, Marjaneh2018, Saadatmand2015, Gani2021}. Such collisions can produce a very large energy density around the collision point, and the properties of the high energy region are important for predicting the maximal number of kink-antikink pairs formed in wave train collisions~\cite{Askari2018}. In some models~\cite{Gani2019, Yan2020}, one can also pinpoint the location of bounce windows by observing the behaviors of maximal energy densities, which vary smoothly within the bounce windows but become chaos elsewhere. Other interesting topics include deformed $\phi^4$ theory~\cite{Adam2020}, deformed sine-Gordon model\cite{Mukhopadhyay2021}, etc.

Despite many interesting discoveries, non-integrable kink collisions are rarely considered in Lorentz-violating models. Considering the fact that Lorentz symmetry breaking naturally arises in many fundamental theories, such as canonical and loop quantum gravity~\cite{AmelinoCamelia1998, Gambini1999, Alfaro2000}, noncommutative quantum field theory~\cite{Anisimov2002}, and might have observational implications~\cite{Colladay1998, Kostelecky2004} (see~\cite{Bietenholz2011} for more references), it would be interesting to consider kink collisions in Lorentz-violating non-integrable scalar field models. However, the breaking of Lorentz symmetry hinders one from deriving moving kink solutions by simply boosting static ones, which are necessary for the construction of the initial configuration. Therefore, the study of kink collisions was limited to the models that preserve the Lorentz invariance for a long time, until Bazeia and others proposed a model in which there is a simple transformation relationship between the static and dynamic kink solutions~\cite{Barreto2006}.

In this paper, we study the kink interactions in a Lorentz-violating $\phi^6$ model using the traveling kink solution reported in Ref.~\cite{Barreto2006}. In the following section, the Lorentz-violating $\phi^6$ model is considered and the kink solutions in the Lorentz-violating $\phi^6$ model are derived. Then we analyze the linear perturbation modes of both the single kink and double kink solutions. The numerical simulation of the kink collisions, as well as the dependencies of several physical quantities on the Lorentz-violating parameter, is conducted subsequently.

\section{Model and solution}

The $(1+1)$-dimensional Lorentz-violating $\phi^6$ theory is defined by the Lagrangian density:
\begin{eqnarray}
\mathcal{L}=\frac{1}{2} \eta^{\mu \nu}\partial_{\mu} \phi \partial_{\nu} \phi+\frac{1}{2} \kappa^{\mu \nu} \partial_{\mu} \phi \partial_{\nu} \phi-\frac{1}{2}\phi^{2}(1-\phi^{2})^{2},
\label{lagrange}
\end{eqnarray}
where we introduced the conventional Minkowski metric $\eta^{\mu \nu}\equiv \begin{pmatrix} 1 & 0 \\ 0 & -1 \end{pmatrix}$ and a tensor $\kappa^{\mu \nu}\equiv \begin{pmatrix} 0 & \alpha \\ \alpha & 0 \end{pmatrix}$, representing the Lorentz violation.

Since the $\phi^6$ model contains three degenerate vacua ${\{-1,0,1\}}$, kinks and antikinks take two types, each of which is connected to neighboring vacua and has to be studied separately.

The Lagrangian (Eq.~\ref{lagrange}) yields the EoM for the field
\begin{equation}
\frac{\partial^{2} \phi}{\partial t^{2}}-\frac{\partial^{2} \phi}{\partial x^{2}}+2 \alpha \frac{\partial^{2} \phi}{\partial x \partial t}+3 \phi^{5}-4 \phi^{3}+\phi=0
\label{eom}
\end{equation}

The static kink solution interpolating $(0, 1)$ is found to be the same as the ordinary $\phi^6$ model solution
\begin{equation}
\phi_K(x) = \phi_{s(0,1)}(x)=\sqrt{\frac{1+\tanh(x)}{2}}.
\label{before_boost}
\end{equation}
The corresponding antikink is
\begin{equation}
\phi_{\bar{K}}(x) = \phi_{(1,0)}(x) = \phi_K(-x),
\end{equation}
and kinks belong to the other topological sector are
\begin{equation}
\phi_{(0,-1)}(x) = - \phi_K(x), \phi_{(-1,0)}(x) = - \phi_K(-x).
\end{equation}

To obtain a moving kink solution, one has to resort to a more sophisticated method, for the ordinary Lorentz boost simply does not hold in this Lorentz-violating model. However, one can still carry out a boost-like approach by deforming the Lorentz factor $\gamma \equiv 1 / \sqrt{1-v^{2}+2 \alpha v}$~\cite{Barreto2006}. The traveling kink solution can then be expressed as
\begin{equation}
\phi_{(0,1)}(x_0, t) = \sqrt{\frac{1+\tanh[\gamma(x-x_0-v_0t)]}{2}},
\label{boost}
\end{equation}
where $x_{0}$ and $v_{0}$ are the initial position and velocity of the kink.

\begin{figure}[h]
\centering
\includegraphics[width=0.49\textwidth]{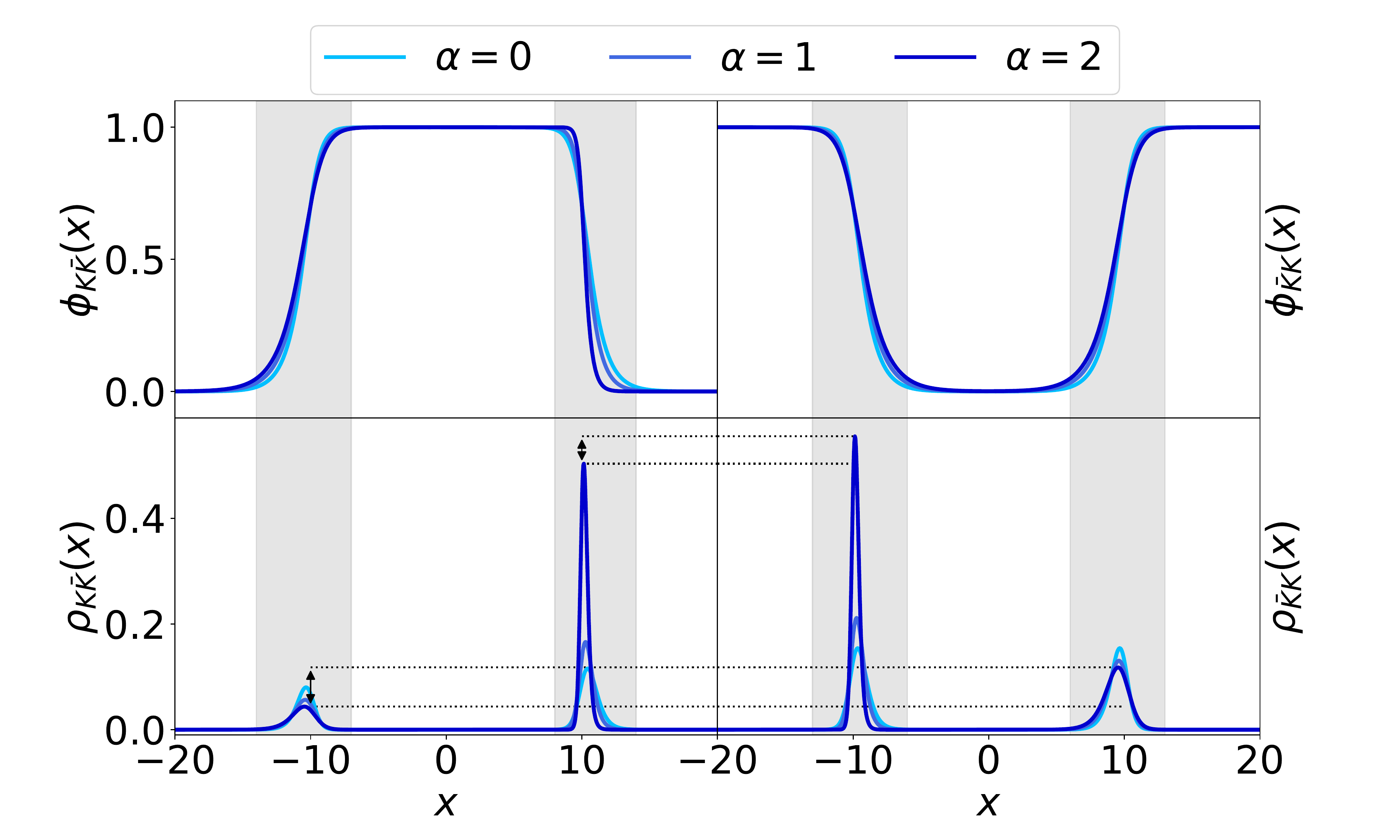}
\caption{Configurations of $K\bar{K}$ and $\bar{K}K$ topology for $v_0=0.2$, $x_0=10$, and $\alpha=0,1,2$. The antikinks are always more energetic than kinks, and this asymmetry grows as $\alpha$ increases. The dashed lines show the difference of maximal energy densities between right-moving and left-moving (anti)kinks.}
\label{fig_kink}
\end{figure}

As is known, the breaking of the Lorentz invariance breaks the symmetry (the width, the energy density, etc) between moving kink and antikink, and kink collisions would be asymmetrical. The spatial configurations of kink-antikink ($K\bar{K}$) and antikink-kink ($\bar{K}K$) pairs at $\alpha=0,1,2$ are shown in Fig.~\ref{fig_kink}. As $\alpha$ increases, the asymmetry between the kink and the antikink increases. Note that the energy density distribution of the same kink is also different if it is moving oppositely, as indicated by the dashed lines.

\section{Linear stability and the vibrational mode}
\label{sec_per}

To prove the stability of the kink solutions and get a better understanding of the bounce windows, the linear perturbative states of the static kink solutions should be examined. Consider a small perturbation $\delta \phi(x,t)$ around a background (a static kink or a two-kink configuration), by taking the variation of the action with respect to $\delta \phi(x,t)$ and Fourier expand the field $\phi(x,t)$, we immediately arrive at a Schr\"odinger-like equation:
\begin{equation}
\mathscr{H} f_n=\tilde{\omega}_n^2 f_n,
\end{equation}
where a factorizable Hamiltonian $\mathscr{H}\equiv -\frac {{\mathrm{d}}^2}{{\mathrm{d}}x^2}+\frac{1}{\theta}\frac {{\mathrm{d}}^2 \theta}{{\mathrm{d}}x^2}=\left(\frac{{\mathrm{d}}}{{\mathrm{d}} x}+\frac{1}{\theta}\frac {{\mathrm{d}} \theta}{{\mathrm{d}}x}\right)\left(-\frac{{\mathrm{d}}}{{\mathrm{d}} x}+\frac{1}{\theta}\frac {{\mathrm{d}} \theta}{{\mathrm{d}}x}\right)$ (with $\theta=\partial_x\phi_s$) is introduced, and $ \tilde{\omega}_n^2\equiv \left(1+\alpha ^2\right) \omega_n^2$ is then always non-negative according to the supersymmetric quantum mechanics, thus proving our claim that the static solutions are stable against small perturbations. Note that $f_{n}(x)$ is only the expansion coefficient of perturbation, and $g_n(x)\equiv f_{n}(x)\mathrm{e}^{{i}\alpha \omega_n x}$ is the complete eigenfunction.

For the single kink solution $\phi_{s(0,1)}(x)=\sqrt{\frac{1+\tanh(x)}{2}}$, the effective potential in the Schr\"odinger-like equation is $V_{\mathrm{eff}}=\frac{\partial_{x}^{3} \phi_{s}}{\partial_{x} \phi_{s}}= \frac{15}{4}(\tanh (x)+1)^{2}-6(\tanh (x)+1)+1$ and only translational mode survives.

\begin{figure*}[h]
\centering
\begin{subfigure}[b]{0.325\textwidth}
\centering
\includegraphics[width=\textwidth]{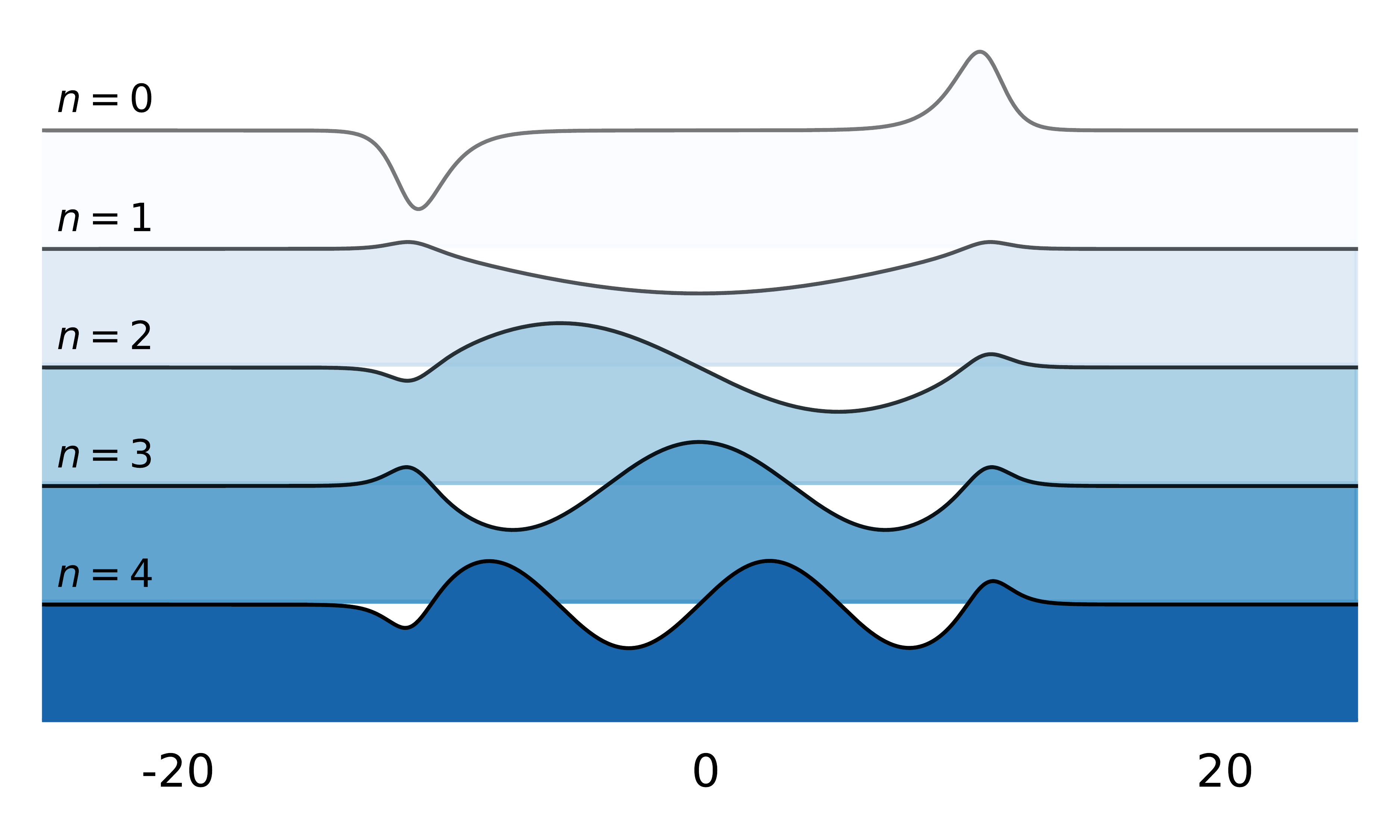}
\caption{$\alpha=0$}
\end{subfigure}
\hfill
\begin{subfigure}[b]{0.325\textwidth}
\centering
\includegraphics[width=\textwidth]{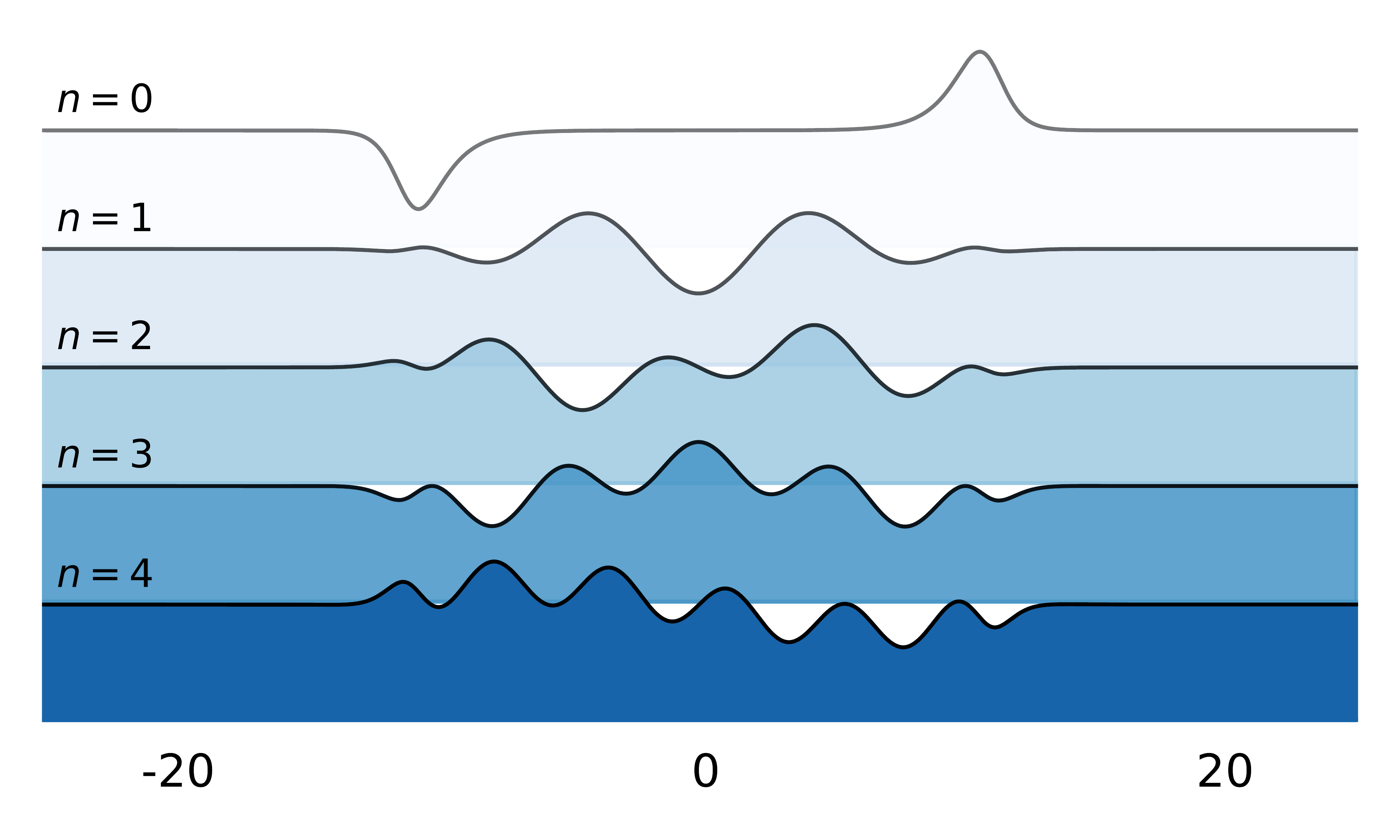}
\caption{$\alpha=1$}
\end{subfigure}
\hfill
\begin{subfigure}[b]{0.325\textwidth}
\centering
\includegraphics[width=\textwidth]{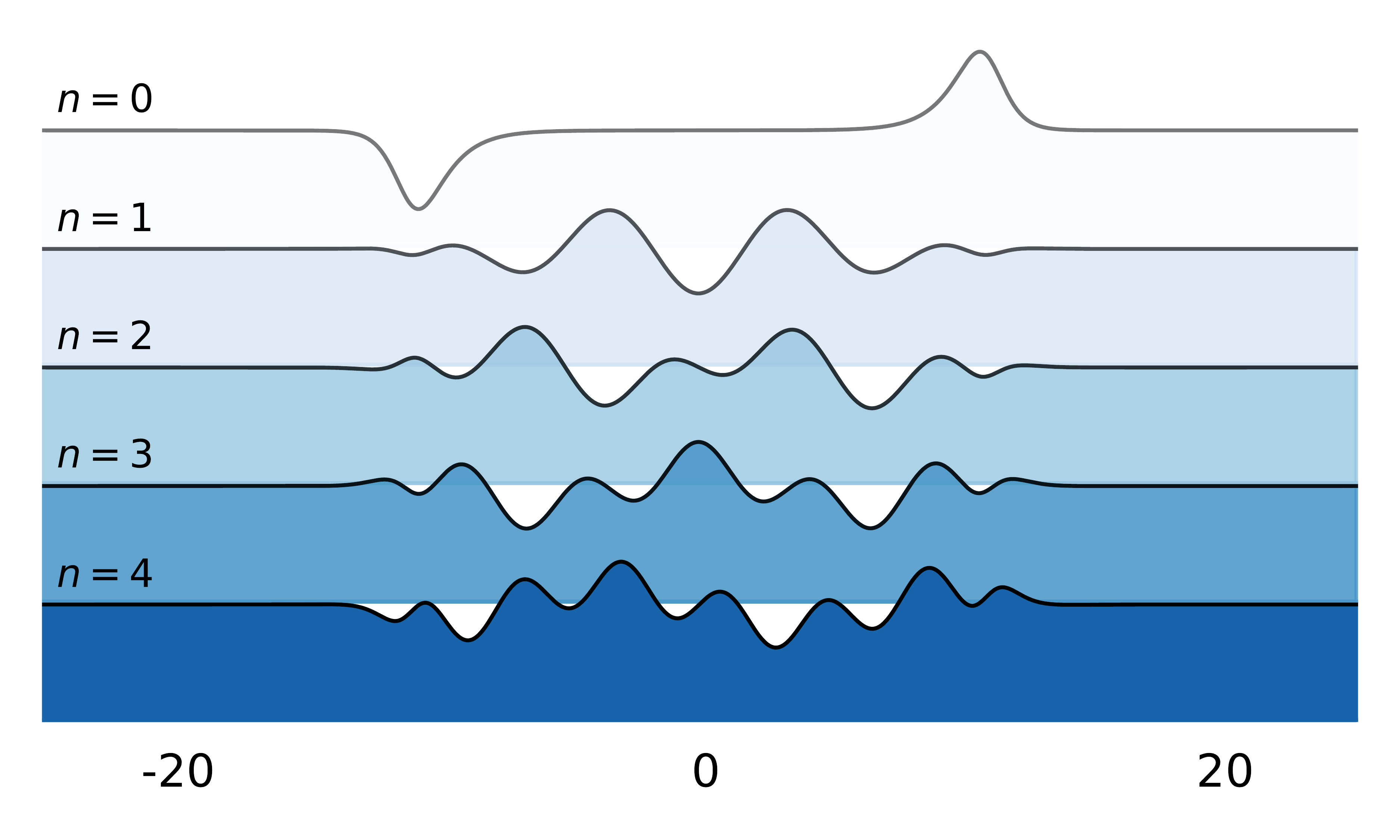}
\caption{$\alpha=2$}
\end{subfigure}
\caption{Eigenfunctions of modes of $\bar{K}K$ static solution for $\alpha=0$ (ordinary $\phi^6$ model) and $\alpha=1, 2$. $0 \leq n \leq 4$ stands for the zero mode and the $n\textsuperscript{th}$ excited state, and $x_0=10$ is the half-separation distance. The spatial wave function $g_n(x)\equiv f_n(x)e^{i \alpha \omega_n x}$ gets more nodes under Lorentz violation.}
\label{fig_waveFunction}
\end{figure*}

For the two-kink static configurations,
\begin{equation}
\begin{cases}
\phi_{K\bar{K}}(x,0)=\phi_{\bar{K}}(x-x_0)+\phi_{K}(x+x_0)-1,\\
\phi_{\bar{K}K}(x,0)=\phi_{K}(x-x_0)+\phi_{\bar{K}}(x+x_0),
\end{cases}
\label{eq:twokink}
\end{equation}
the effective potential becomes $V_{\mathrm{eff}}^{2K}=
15\phi_{2K}^{4}-12\phi_{2K}^{2}+1$, where $\phi_{2K}$ denotes either $K\bar{K}$ or $\bar{K}K$ pair. It turns out that the $K\bar{K}$ pair still supports only translational mode, but $\bar{K}K$ configuration supports many of the vibrational modes, just as the case in the ordinary $\phi^6$ model~\cite{Dorey2011}. The eigenvalues $\tilde{\omega}_n^2$ depend both on the half-separation $x_0$ and the Lorentz-violating parameter $\alpha$. The eigenfunctions $g_{n}(x)$ for the first five modes of $\bar{K}K$ pair with half-separation $x_0=10$ ($14$ modes in total) are shown in Fig.~\ref{fig_waveFunction}. The $\alpha-$dependence of the vibrational modes indicates that the kink collision results will be affected by the Lorentz violation. Various novel kink-collision phenomena will be discussed in the following sections.

\section{Simulation and results}
\label{sec_sim}

In this section, we study the asymmetrical kink collisions. Due to the absence of exact kink scattering solutions of the non-integrable model, one has to resort to numerical simulation. The initial conditions of kink scattering are set by superpositions of individual kinks and antikinks, which is Eq.~\ref{eq:twokink} and

\begin{equation}
\begin{cases}
\dot{\phi}_{K\bar{K}}(x,0)=\dot{\phi}_{\bar{K}}(x-x_0)+\dot{\phi}_{K}(x+x_0),\\
\dot{\phi}_{\bar{K}K}(x,0)=\dot{\phi}_{K}(x-x_0)+\dot{\phi}_{\bar{K}}(x+x_0).
\end{cases}
\end{equation}

Since a single kink has an exponentially decaying tail, the overlap in the initial conditions can be safely ignored at a large distance. Initial conditions for more kinks can be obtained in the same way.

The Fourier spectrum method~\cite{Trefethen2000} with periodic boundary conditions is used to solve the EoM numerically, and one may control the validity of the simulation by insisting on energy conservation. The total energy of the configuration is calculated by
\begin{equation}
E(t)=\frac{1}{2} \int_{-\infty}^{\infty}\left[\left(\frac{\partial \phi}{\partial t}\right)^{2}+\left(\frac{\partial \phi}{\partial x}\right)^{2}+\frac{1}{2} \phi^{2}\left(1-\phi^{2}\right)^{2}
\right] \mathrm{d} x,
\end{equation}
and can be approximated by
\begin{equation}
E_{\mathrm{th}} \approx E\left(v_{0}\right)+E\left(-v_{0}\right),
\end{equation}
where $E\left(v_{0}\right)=\gamma\left(1+\alpha v_{0}\right) M_K$ is the energy of a moving kink and the mass of the kink is $M_K=\int[(\partial_{x} \phi_{(0,1)})^{2} / 2+V] d x=\frac{1}{4}$. The relative error between $E_{\mathrm{th}}$ and the calculated energy $E_{\mathrm{num}}$ will be checked to guarantee the validity of simulation results. The EoM is discretized, with the time step chosen automatically by MATLAB ode45 solver, typically around $0.002$, to ensure the relative error of energy to be less than $10^{-9}$. The space step is chosen to be $0.1$ and $0.05$ to guarantee the results are convergent and no significant change happens with further precision improvement.

\subsection{Impacts on the collision structure}

\begin{figure}[h]
\centering
\begin{subfigure}[b]{0.49\textwidth}
\centering
\includegraphics[width=\textwidth]{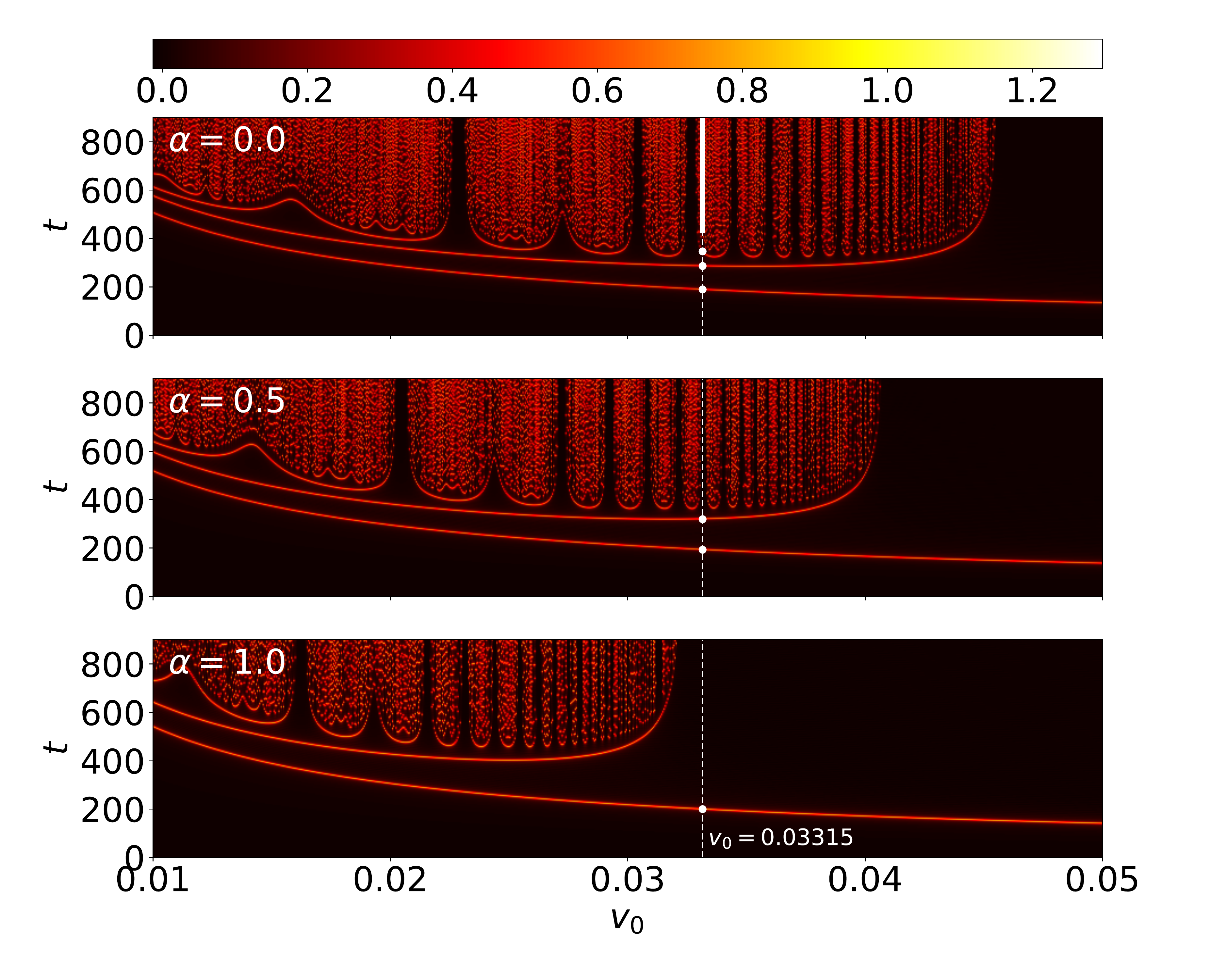}
\caption{$\bar{K}K$}
\label{fig:kbark}
\end{subfigure}
\hfill
\begin{subfigure}[b]{0.49\textwidth}
\centering
\includegraphics[width=\textwidth]{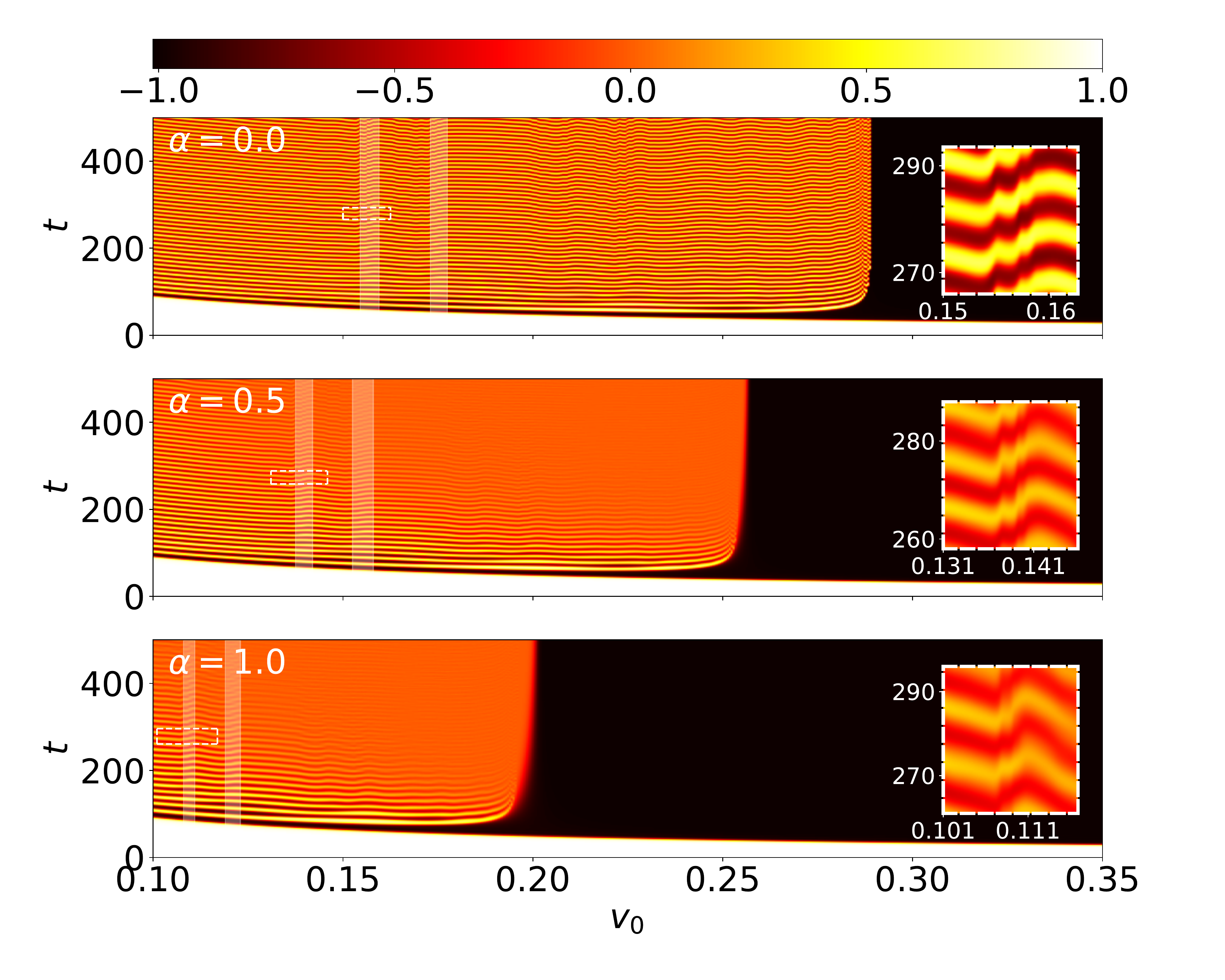}
\caption{$K\bar{K}$}
\label{fig:kkbar}
\end{subfigure}
\caption{Field values of collisions of (a) $\bar{K}K$ and (b) $K\bar{K}$ at the central spatial point as functions of velocities for $\alpha=0, 0.5$ and $1$. As $\alpha$ increases, the critical velocity decreases. There is no 3BW and higher-order bounce window in this figure. Collision points at initial velocities $v_0=0.03315$ have been colored white in plot (a). In plot (b), there is no BW, but there exist windows (RW) where the $n\textsuperscript{th}$ oscillation moments increases as $v_0$ increases, with the first two RWs highlighted and the first RW enlarged on the right.}
\label{fig_structure}
\end{figure}

It is important to look at how the Lorentz-violating term affects the global distribution of BW and $v_c$. The $K\bar{K}$ and $\bar{K}K$-type collision structures for $\alpha=0,0.5,1$ are shown in Fig.~\ref{fig_structure}. As $\alpha$ increases, the critical velocity decreases. The critical difference between this structure and that of the $\phi^4$ model is no bounce window is found in $K\bar{K}$ collisions, and no bounce window higher than the $2\textsuperscript{nd}$ order in $\bar{K}K$ collisions is found, rendering the structure of the Lorentz-violating $\phi^6$ model terminated after the second level, which is consistent with the ordinary $\phi^6$ model~\cite{Dorey2011}.

\begin{figure}[h]
\centering
\includegraphics[width=0.49\textwidth]{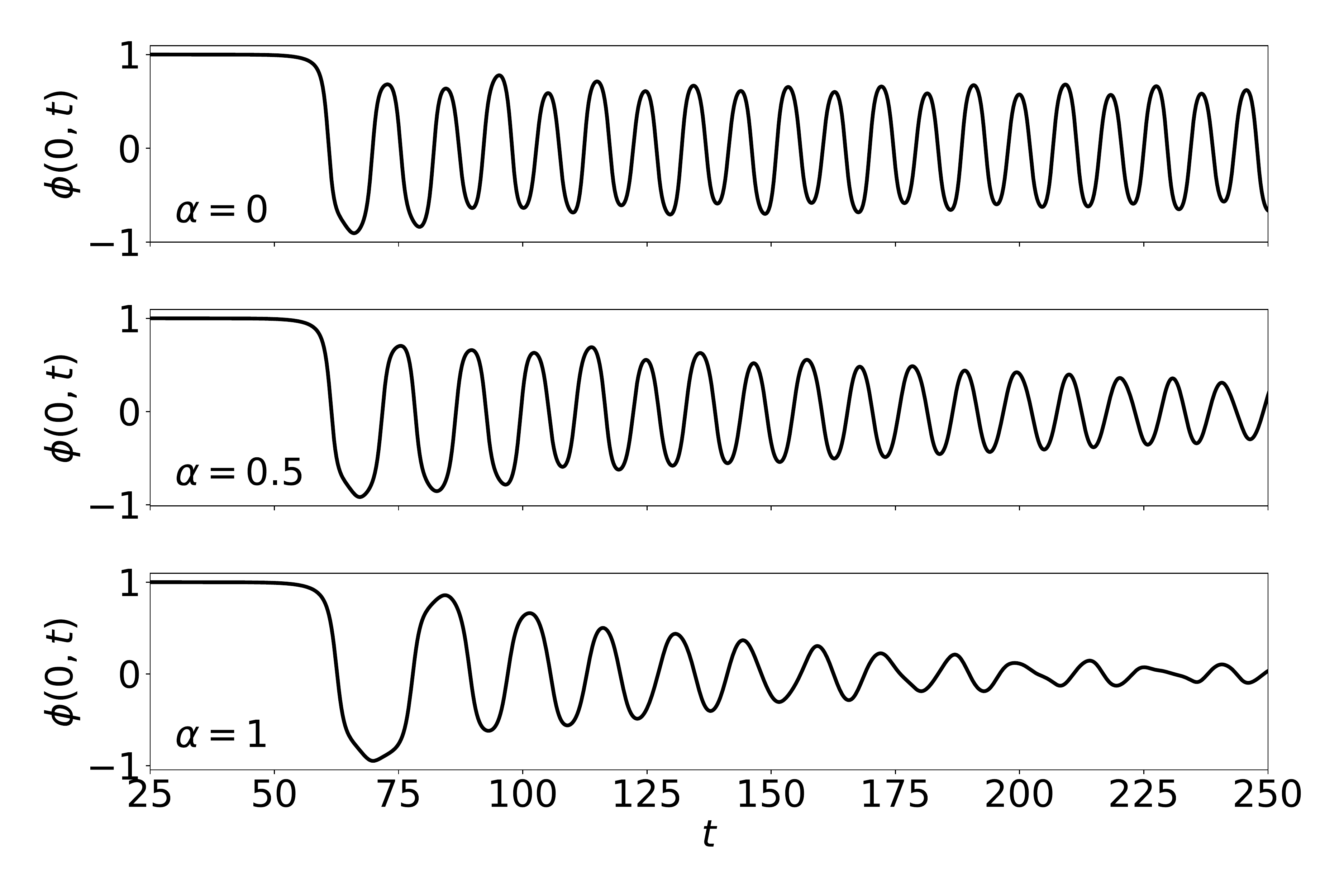}
\caption{The field values at the origin of $v_0=0.15$ for $\alpha=0,0.5,1$ of $K\bar{K}$ collisions. The amplitudes of oscillations decay if $\alpha \neq 0$.}
\label{fig:damping}
\end{figure}

One can make another two observations regarding the oscillations in the $K\bar{K}$ structure, as shown in Fig.~\ref{fig:kkbar}. First, the moments of $n\textsuperscript{th}$ oscillations of the bions plot a series of curves as functions of $v_0$. As $v_0$ increases, the curves decrease in general, but there exist certain windows where all the curves start to rise, which is called the rising windows (RW). The existence of RWs is also confirmed in Lorentz-violating collisions. Furthermore, by zooming in on the windows one would find the RWs decompose into smaller increasing and decreasing segments. The first two RWs are highlighted, and the first is enlarged on the right. Second, in the ordinary $\phi^6$ model, the amplitudes of the oscillations are relatively preserved, while the presence of Lorentz violation damps the oscillation. This damping effect is most evident near the critical velocity, where the amplitudes of oscillations quickly decay to near zero. The field values at the origin of $v_0=0.15$ for $\alpha=0,0.5,1$ are plotted in Fig.~\ref{fig:damping} to illustrate the damping effect. We emphasize that this is due to the asymmetry of the Lorentz-violating $K\bar{K}$ collisions. The bions formed from the collisions deviate from $x=0$ as time evolves, so the field values at $x=0$ come mainly from radiations instead of the bions themselves. However, the bions formed in the $\bar{K}K$ collisions are not found to deviate from the collision points.

\begin{figure*}[h]
\centering
\begin{subfigure}[b]{0.325\textwidth}
\centering
\includegraphics[width=\textwidth]{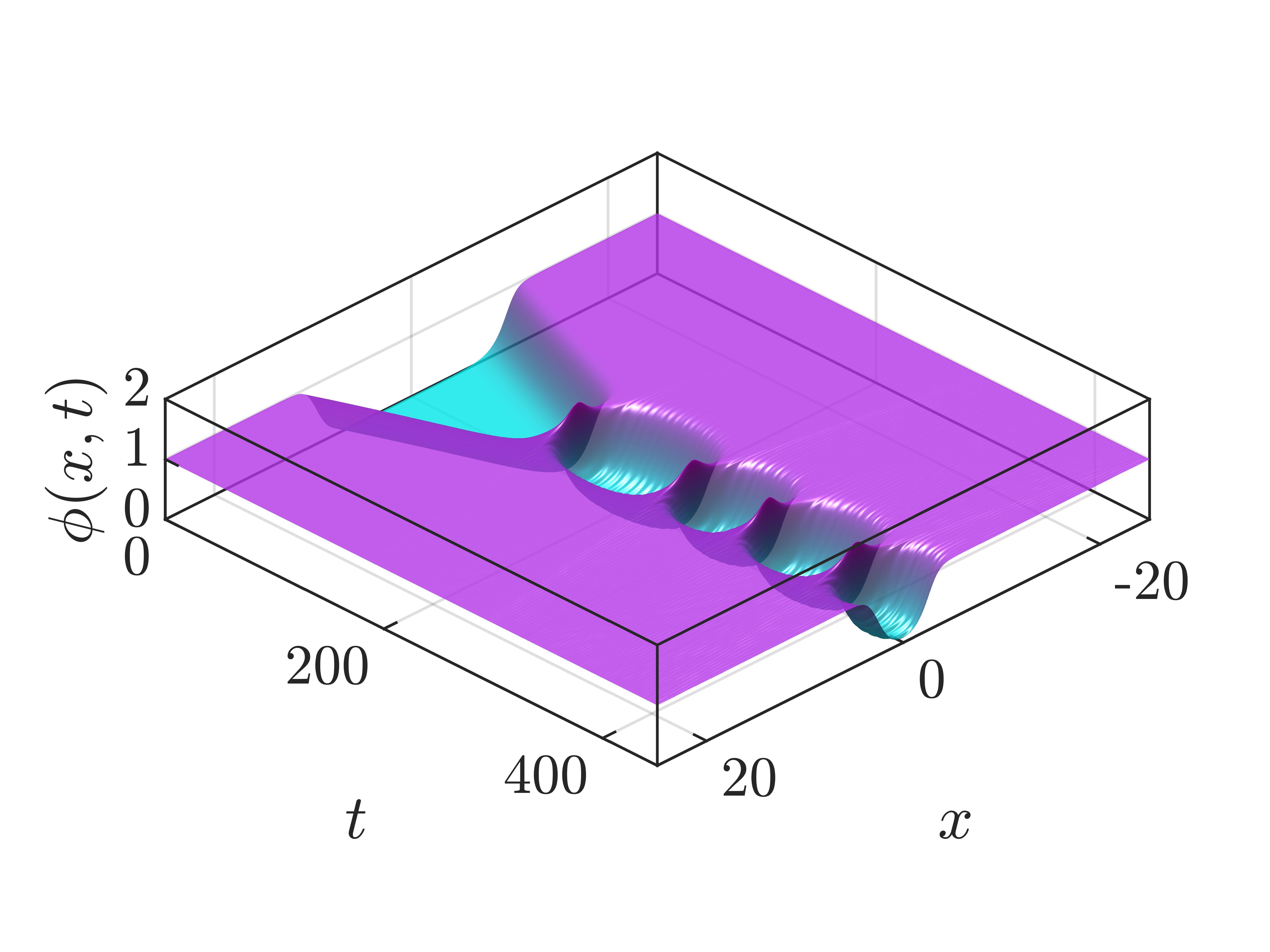}
\end{subfigure}
\hfill
\begin{subfigure}[b]{0.325\textwidth}
\centering
\includegraphics[width=\textwidth]{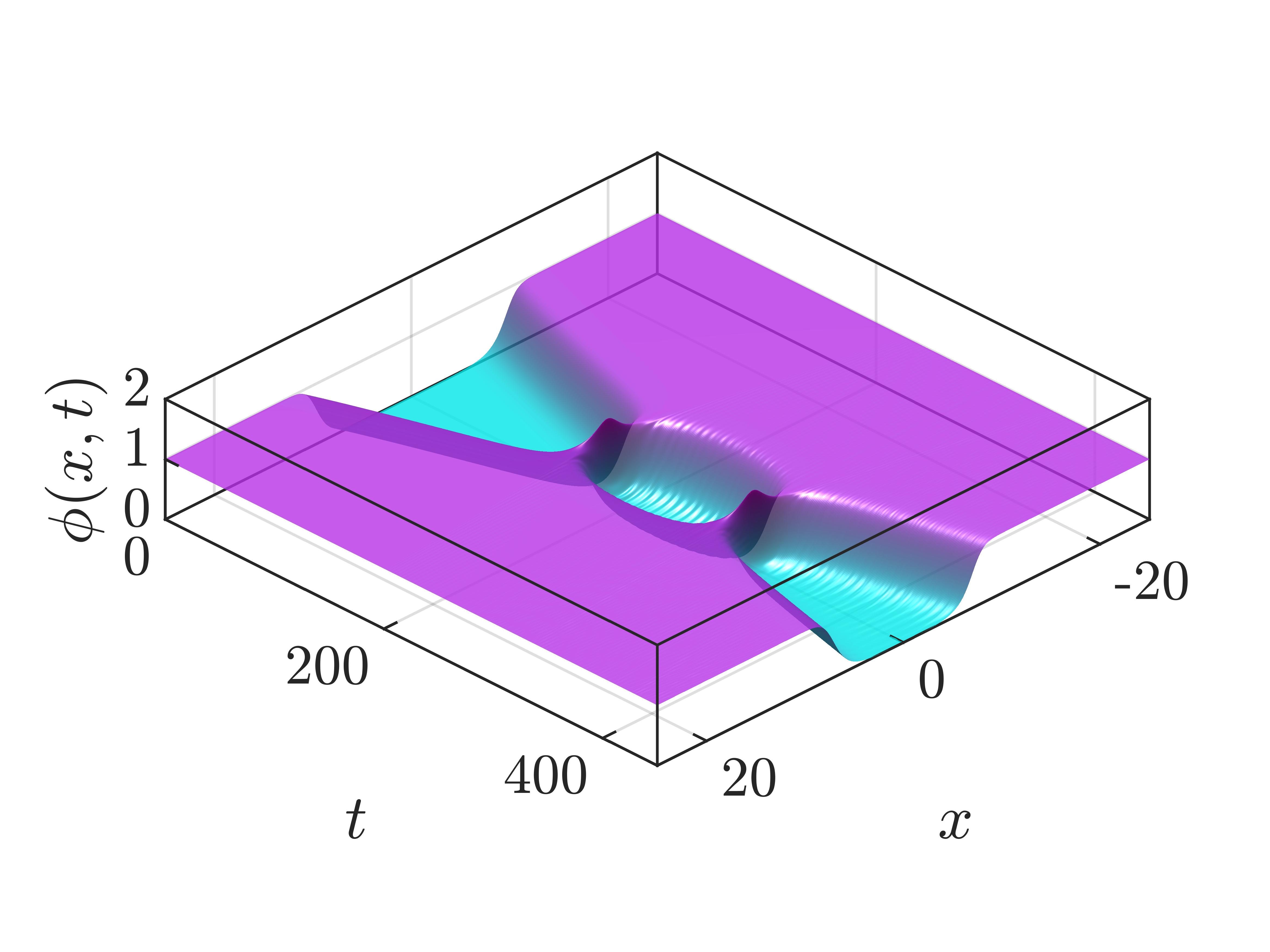}
\end{subfigure}
\hfill
\begin{subfigure}[b]{0.325\textwidth}
\centering
\includegraphics[width=\textwidth]{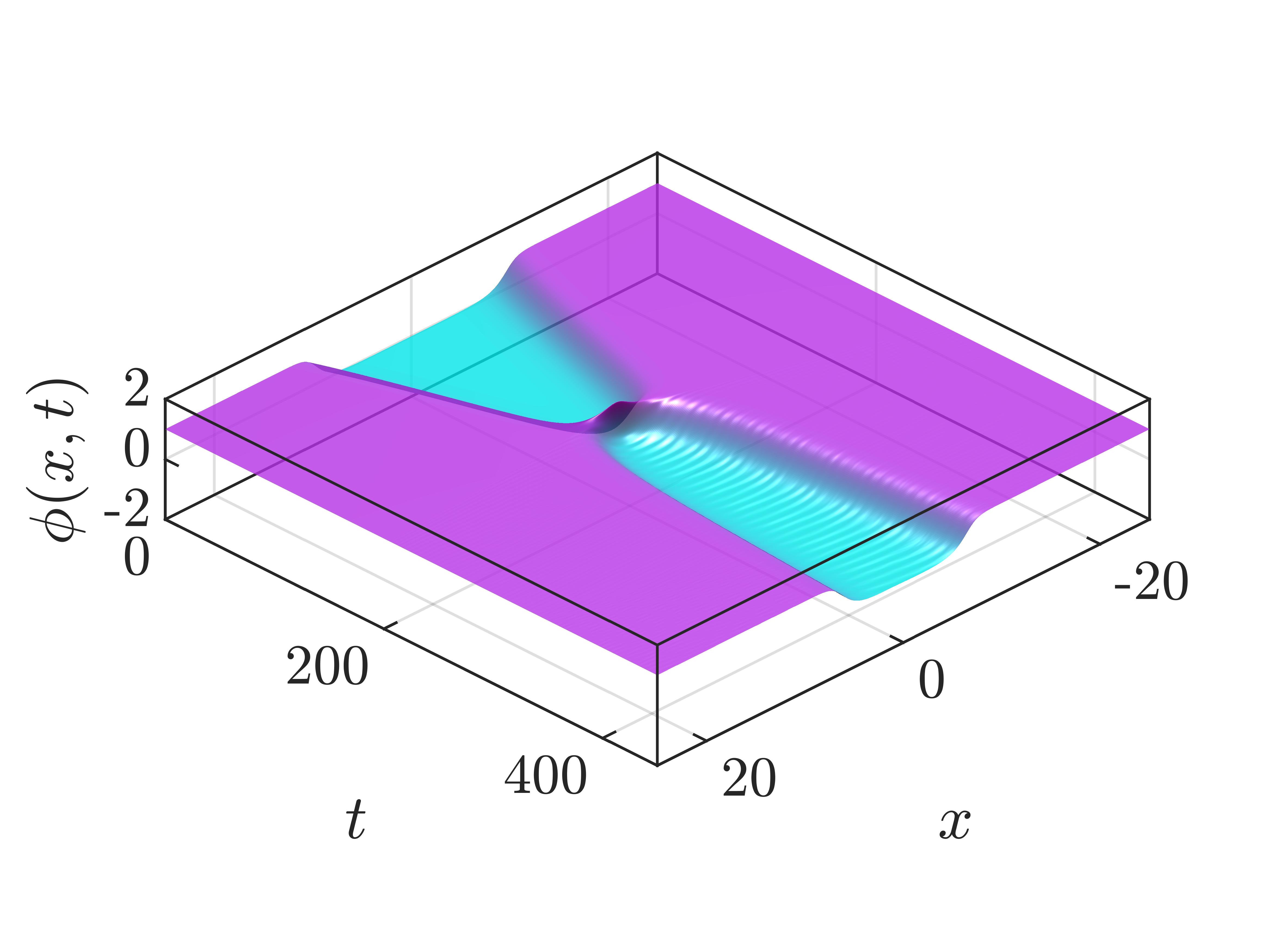}
\end{subfigure}
\quad
\begin{subfigure}[b]{0.325\textwidth}
\centering
\includegraphics[width=\textwidth]{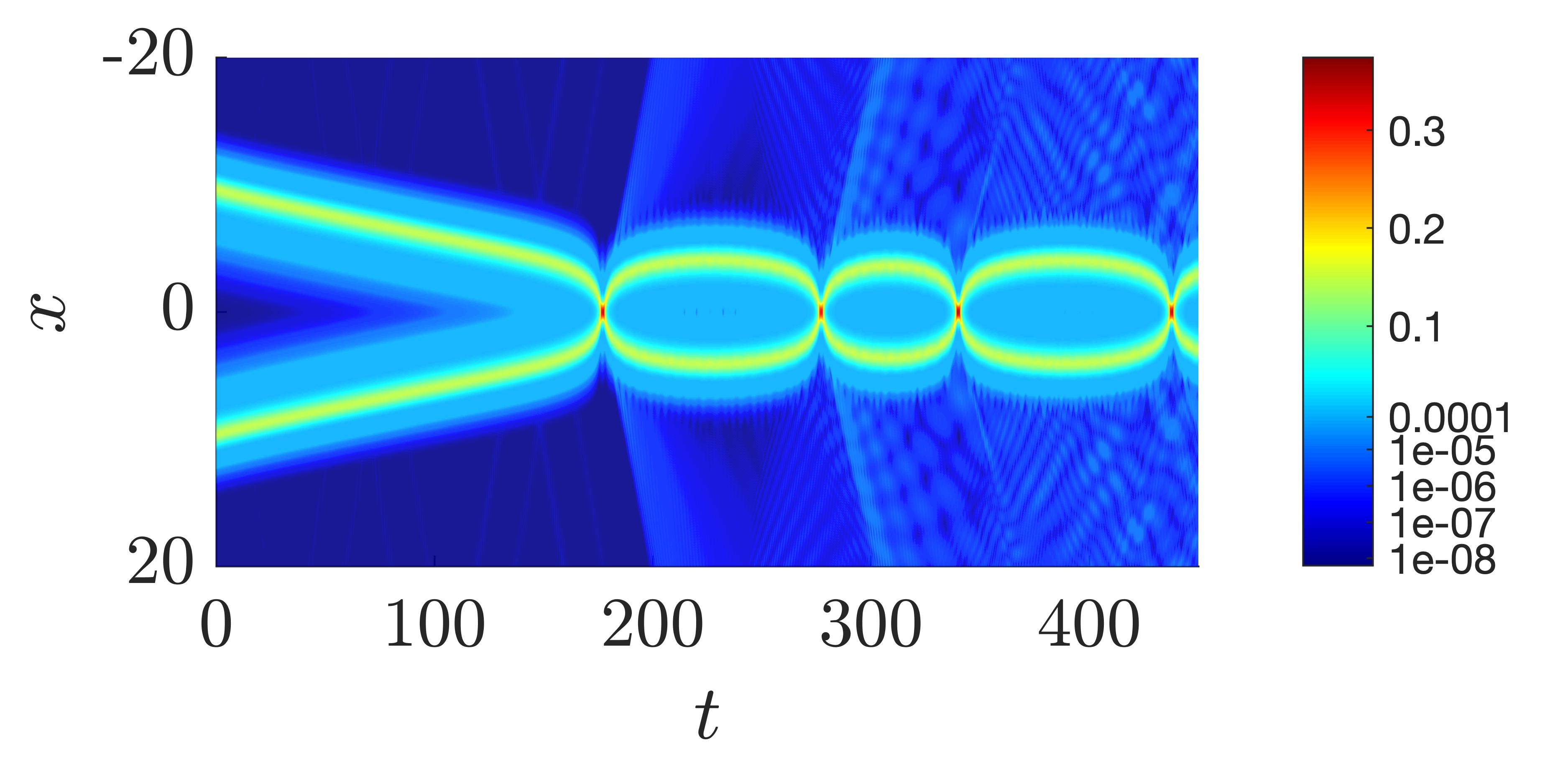}
\caption{$\alpha=0$, bion.}
\label{larger_vc}
\end{subfigure}
\hfill
\begin{subfigure}[b]{0.325\textwidth}
\centering
\includegraphics[width=\textwidth]{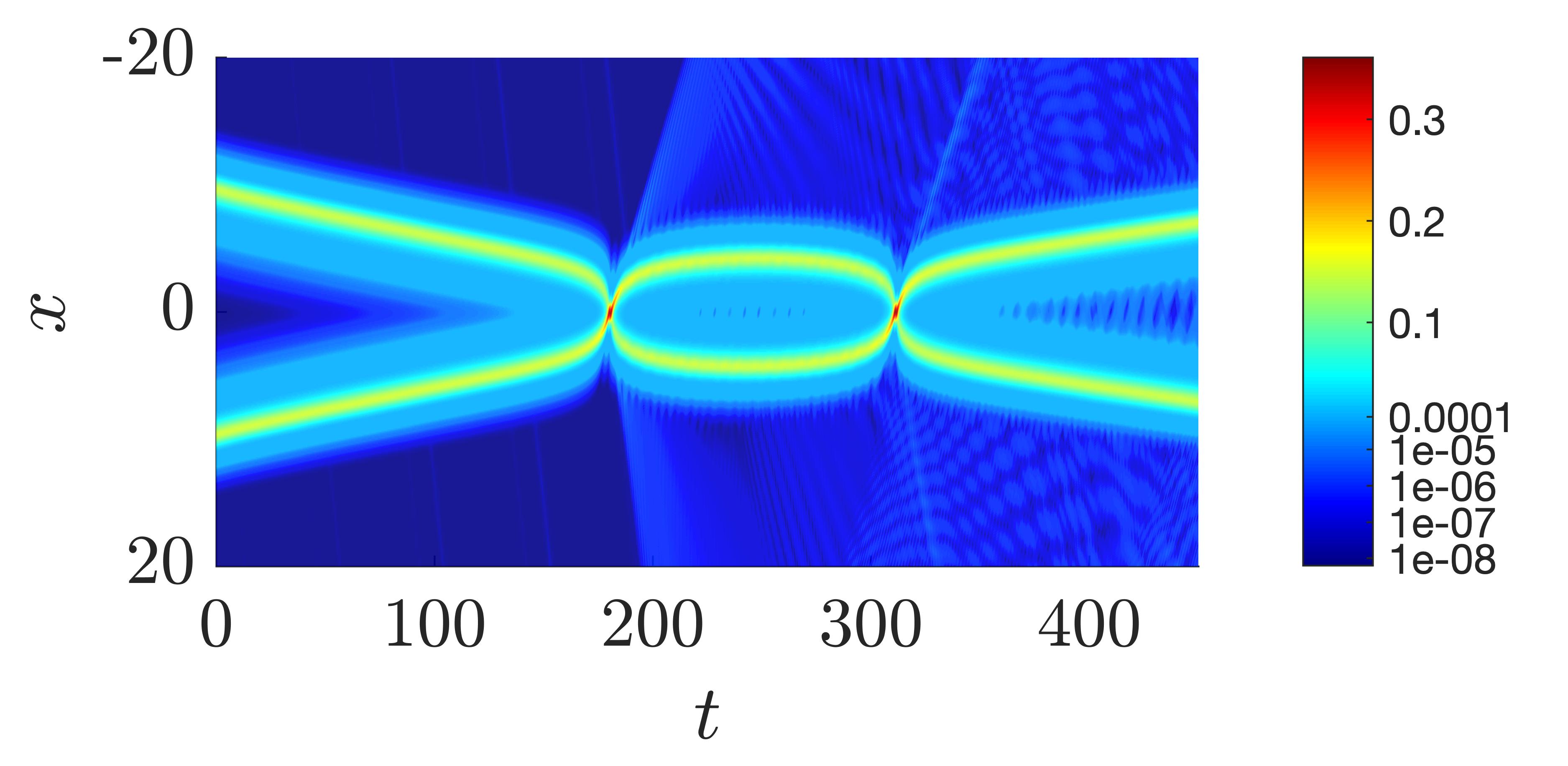}
\caption{$\alpha=0.5$, two-bounce.}
\label{less_vc}
\end{subfigure}
\hfill
\begin{subfigure}[b]{0.325\textwidth}
\centering
\includegraphics[width=\textwidth]{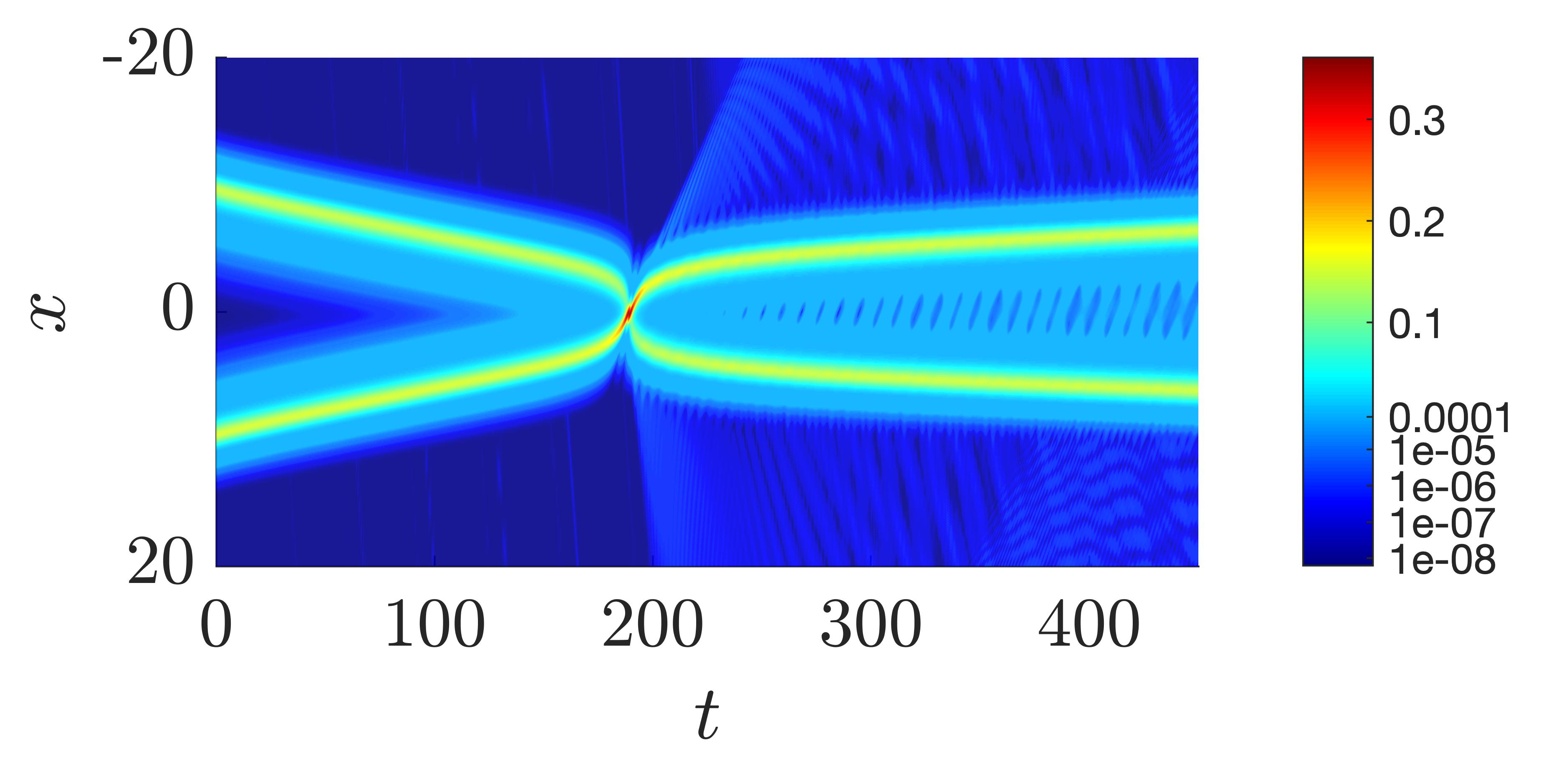}
\caption{$\alpha=1$, inelastic scattering.}
\label{tbw_collision}
\end{subfigure}
\caption{The numerical results for antikink-kink collision at $v_0=0.03315$. The first row shows the field configuration $\phi(x,t)$, and the second row shows the energy density $\rho(x,t)$. The left, middle and right columns are results of $\alpha=0$, $\alpha=0.5$, and $\alpha=1$, and correspond to the bound state, two-bounce window, and inelastic scattering solutions, respectively.}
\label{fig_collision}
\end{figure*}

\begin{figure}[h]
\centering
\includegraphics[width=0.49\textwidth]{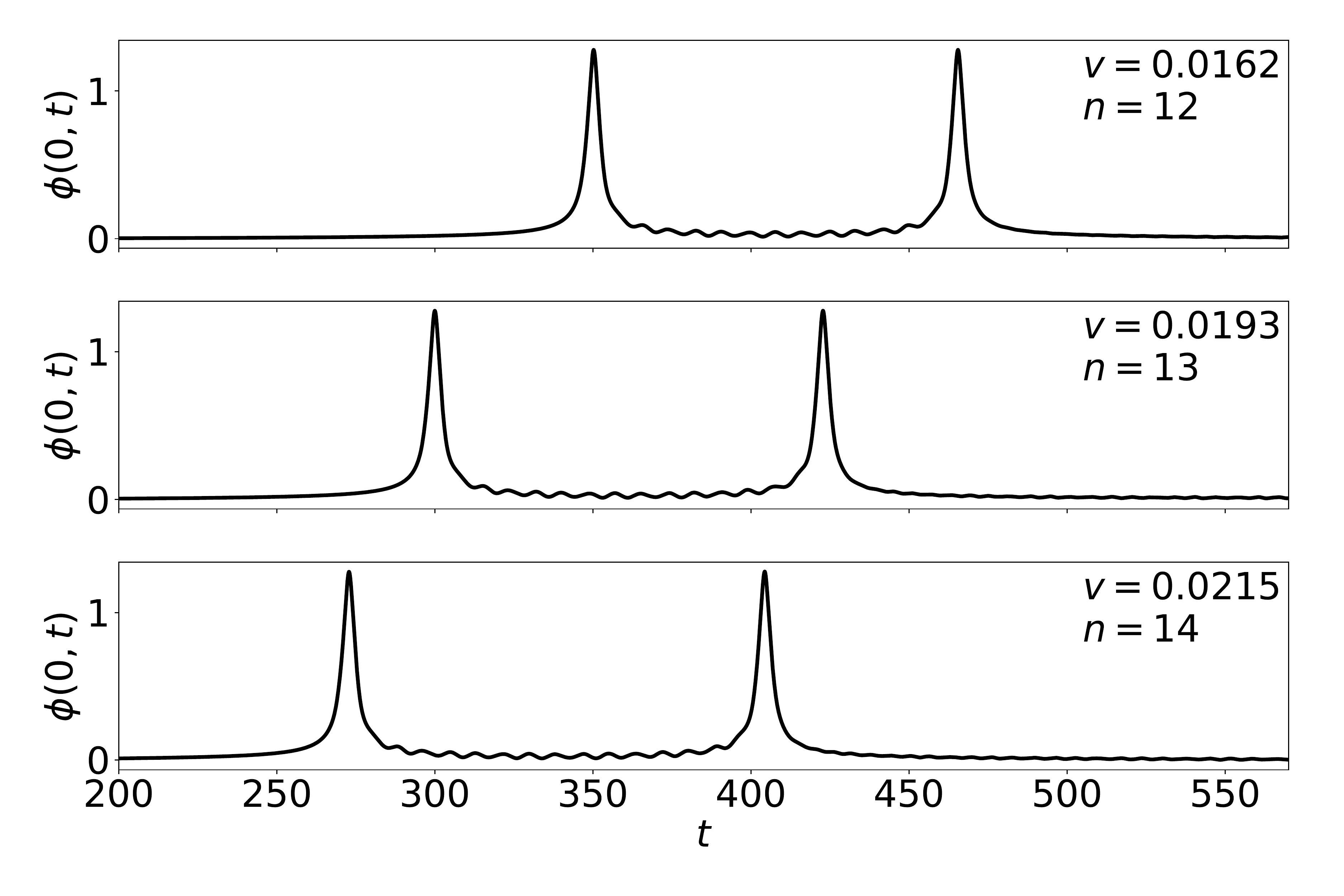}
\caption{The field values at the origin of the first three 2BWs (including a false window) of $\bar{K}K$ collisions at $\alpha=1$. The numbers of oscillations are the same as those at $\alpha=0$.}
\label{fig_phi0t}
\end{figure}

The typical numerical results with $v_0=0.03315$ for antikink-kink collisions have been selected in Fig.~\ref{fig:kbark} and shown in Fig.~\ref{fig_collision}. The first row of figures shows all 3 types of $\phi(x,t)$ configurations, namely bion, two-bounce, and inelastic scattering. The second row shows the corresponding energy density configurations, where the asymmetrical nature of collisions is clearly shown by the behavior of radiation. Note that albeit the asymmetry, the scattering still happens at $x=0$, which can be seen in this row of energy density plots and in Fig.~\ref{fig_structure} where the amplitudes of the first oscillation at $x=0$ are almost the same for $\alpha=0,0.5$, and $1$. The field values at the origin of the first three 2BWs (with the second 2BW being a false window) at $\alpha=1$ are presented in Fig.~\ref{fig_phi0t}. The number $n$ of oscillations of $\phi(0,t)$ between two collisions is the same as that of the ordinary $\phi^6$ model~\cite{Dorey2011}, which together with Fig.~\ref{fig_structure} means that the Lorentz violation may change the positions of 2BWs and $v_c$, but preserve certain invariant quantities, i.e. the number of collective mode oscillations and the number of 2BWs.

\begin{figure}[h]
\centering
\includegraphics[width=0.49\textwidth]{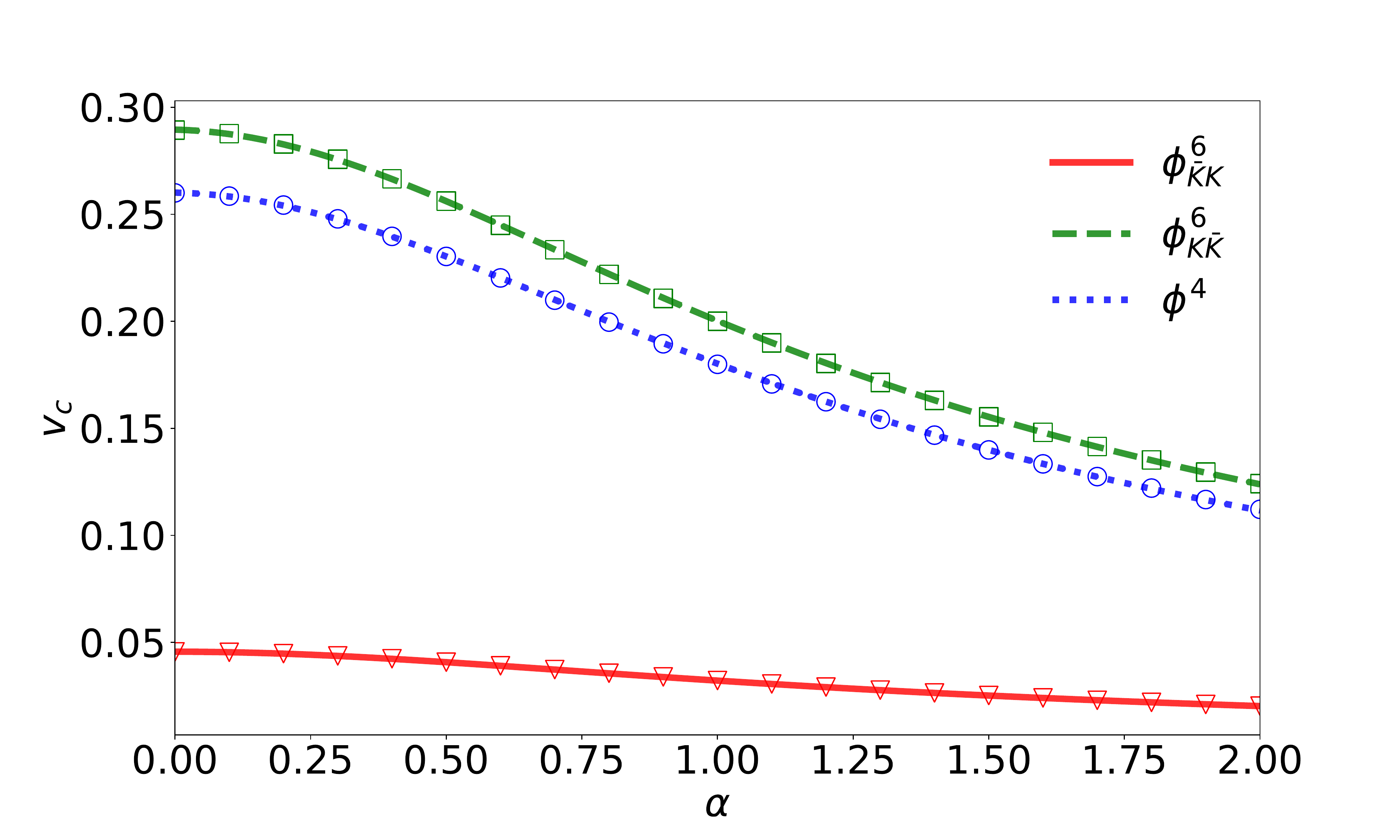}
\caption{The critical velocity $v_{c}$ of $\phi^4$ and $\phi^6$ model collisions and reconstructed fitting curves as functions of the Lorentz-violating parameter $\alpha\in[0, 2]$. The fitted decaying parameters $\tau$ are $0.046, 0.290, 0.260$ for $\bar{K}K$ collision, $K\bar{K}$ collision, and $\phi^4$ collision, respectively.}
\label{fig_fit}
\end{figure}

The critical velocity of $K\bar{K}$ and $\bar{K}K$ collisions in the Lorentz-violating $\phi^6$ model are both monotonically decreasing functions of the parameter $\alpha$, as shown in the data points of Fig.~\ref{fig_fit}. This is just the case in the $\phi^4$ model. And it is of great interest if these results are compared with those of the Lorentz-violating $\phi^4$ model. Despite being calculated from different scalar models or different topologies, the data of the critical velocities present similar varying trends. The following empirical formula is introduced to fit the data:
\begin{equation}
v(\alpha)=v(0)\frac{\mathrm{e}^{-\tau \alpha}}{\sqrt{1+\alpha^2}},
\label{eq_fit}
\end{equation}
and the corresponding fitted curves are also shown in Fig.~\ref{fig_fit}.

To explain their similar behavior under Lorentz invariance breaking, notice that the energies of the kink and the antikink are no longer the same (see Fig.~\ref{fig_kink}). The excessive energy of the kink (antikink) is then wasted, as a result, higher $\alpha$ means lower effective colliding velocity. Eq.~\ref{eq_fit} then provides a prediction for the critical velocities under Lorentz violation.

The dependence of the maximal energy densities on $\alpha$ provides rich information about the kink-scattering behavior under Lorentz breaking. The total energy density can be written as
\begin{equation}
\rho(x,t)=k(x,t)+u(x,t)+p(x,t),
\end{equation}
where $k(x,t)=\frac{1}{2}\left(\frac{\partial \phi}{\partial t}\right)^{2}$, $u(x,t)=\frac{1}{2}\left(\frac{\partial \phi}{\partial x}\right)^{2}$,
and $p(x,t)=V(\phi)$ are the kinetic energy density, gradient energy density, and potential energy density, respectively.

\begin{figure}[h]
\centering
\includegraphics[width=0.49\textwidth]{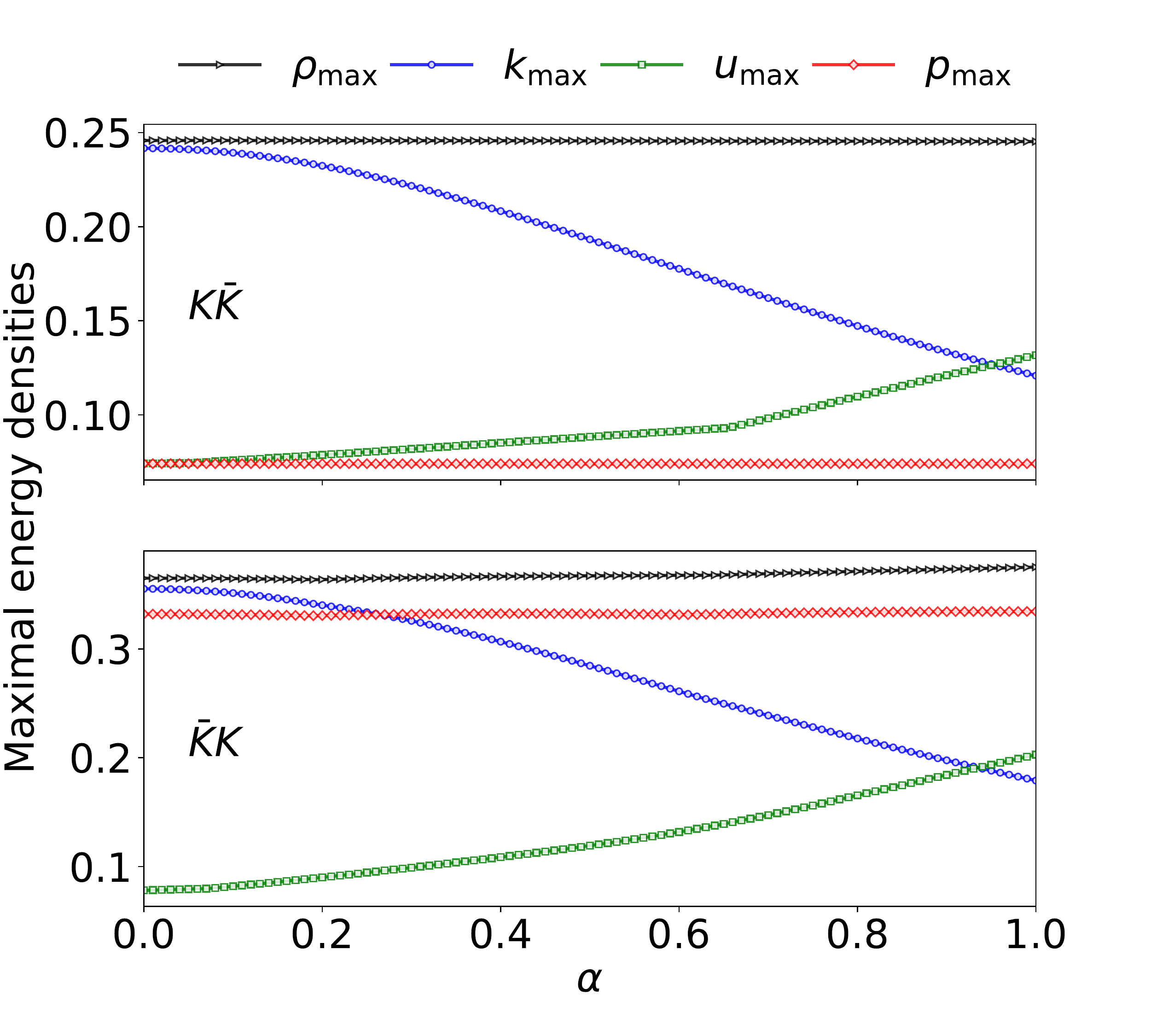}
\caption{Maximal values of $\rho(x,t)$ and its components for $v_0=0.03315$ kink scattering as functions of $\alpha$.}
\label{fig_max}
\end{figure}

From the experience of Ref.~\cite{Gani2019, Yan2020}, one would expect this energy density structure to be chaotic unless the velocity falls in the inelastic scattering region or the bounce windows, showing fractal information about the scattering structure. However, this is not the case anymore for the $\phi^6$ model. In Fig.~\ref{fig_max}, maximal values of $\rho(x,t)$ and its components for kink scattering, with initial conditions $x_0=10$ and $v_0=0.03315$, are presented. The curve is smooth and no chaotic zone is observed, regardless of the final state of the configuration. The lacking of the single vibrational mode forbids the violent energy exchange between different energy levels, the maximal energy densities then lost their fractal chaotic behavior.

\subsection{Impacts on multi-kink collision}
\label{sec_multi}

\begin{figure*}[h]
\centering
\begin{subfigure}[b]{0.325\textwidth}
\centering
\includegraphics[width=\textwidth]{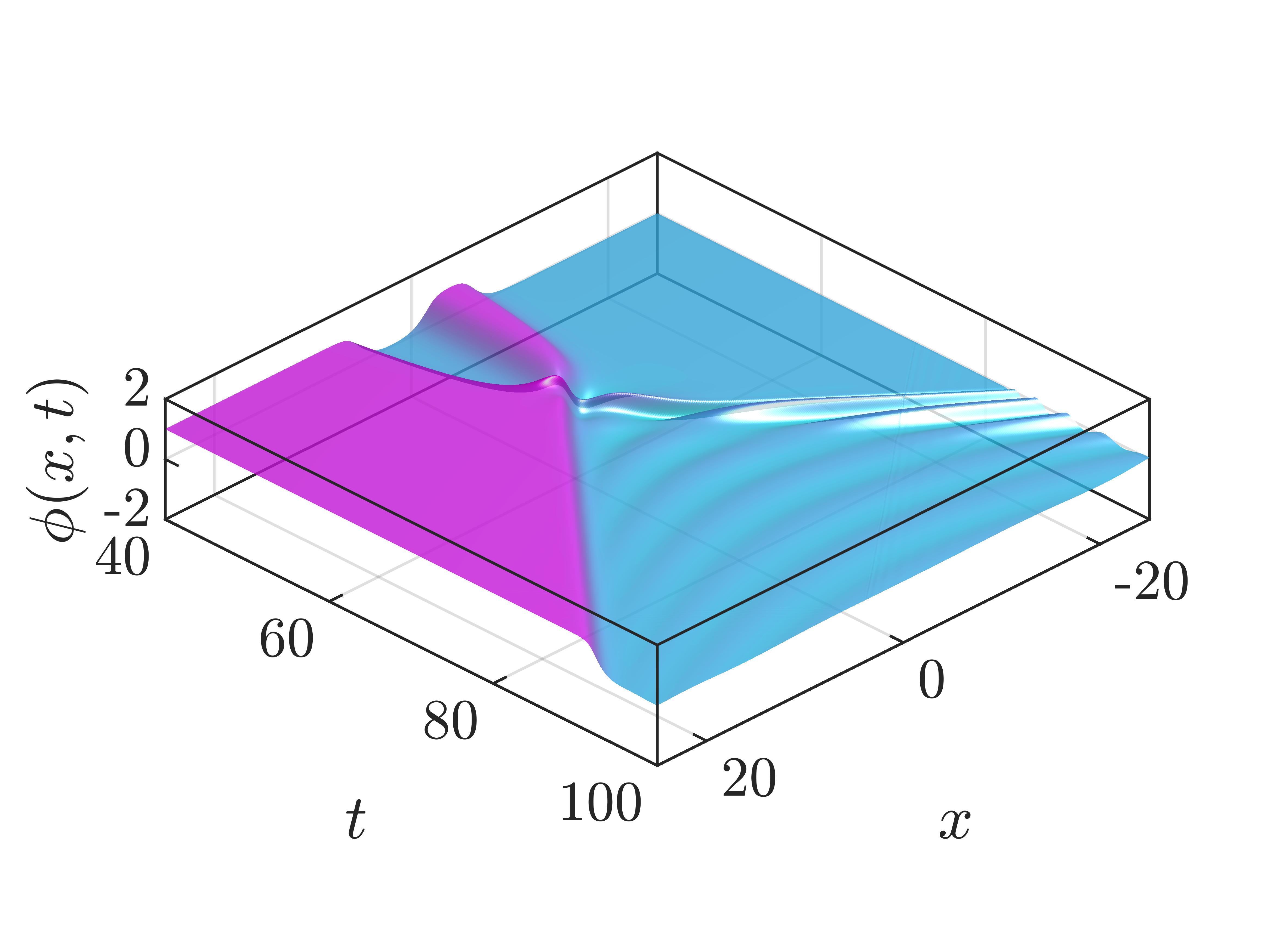}
\end{subfigure}
\hfill
\begin{subfigure}[b]{0.325\textwidth}
\centering
\includegraphics[width=\textwidth]{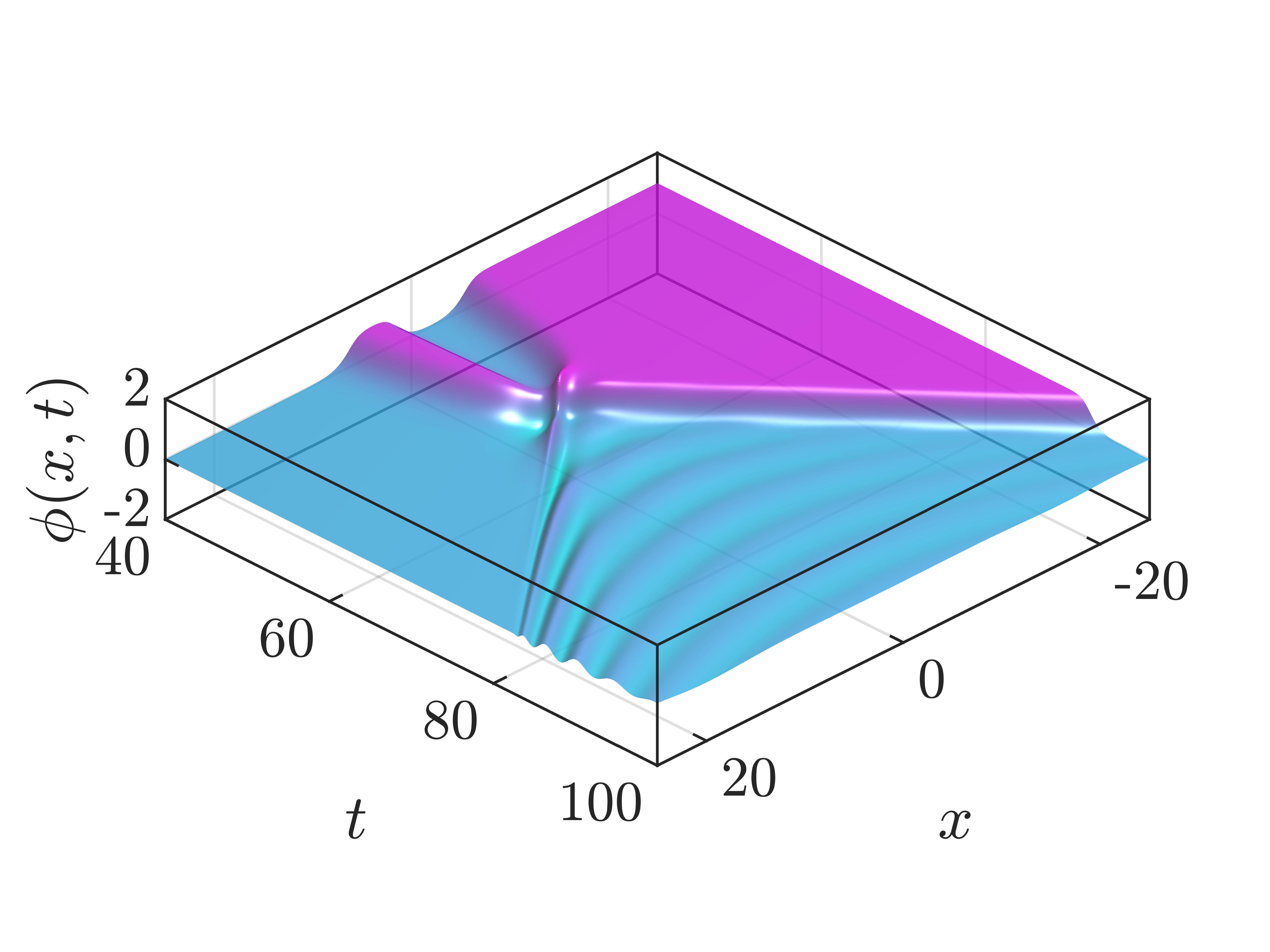}
\end{subfigure}
\hfill
\begin{subfigure}[b]{0.325\textwidth}
\centering
\includegraphics[width=\textwidth]{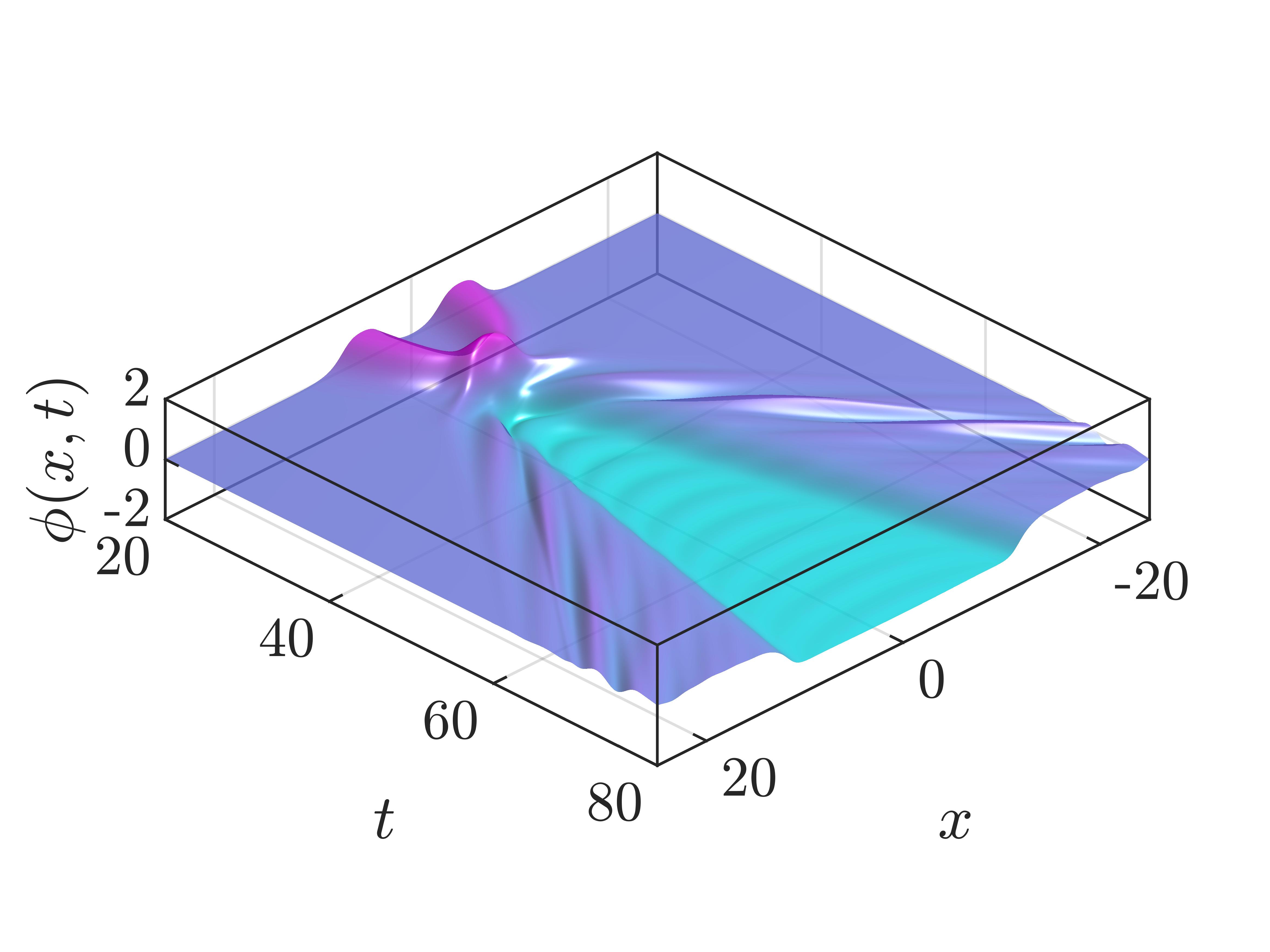}
\end{subfigure}
\quad
\begin{subfigure}[b]{0.325\textwidth}
\centering
\includegraphics[width=\textwidth]{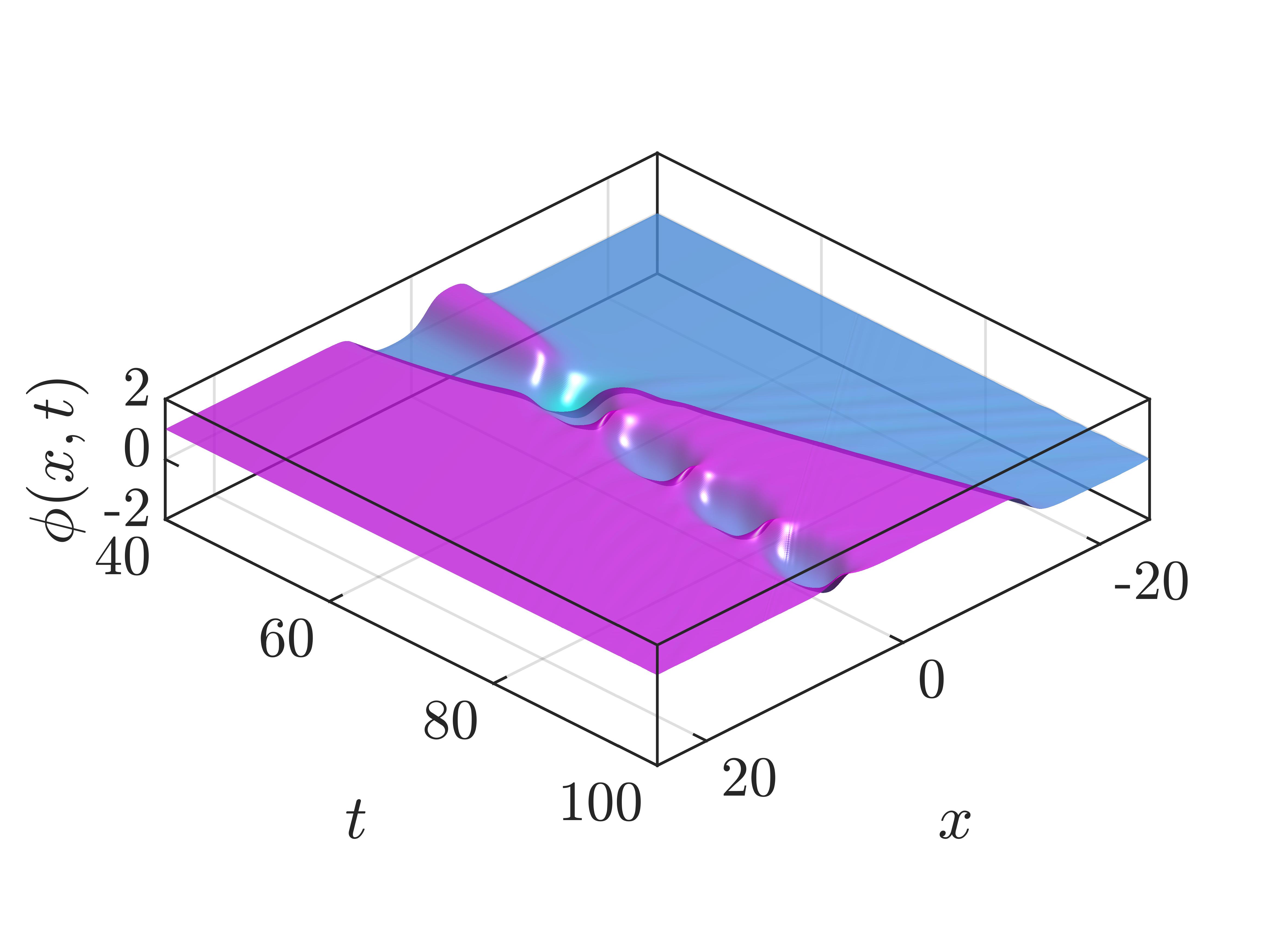}
\caption{$(0,1,0,1)$}
\label{fig_0101}
\end{subfigure}
\hfill
\begin{subfigure}[b]{0.325\textwidth}
\centering
\includegraphics[width=\textwidth]{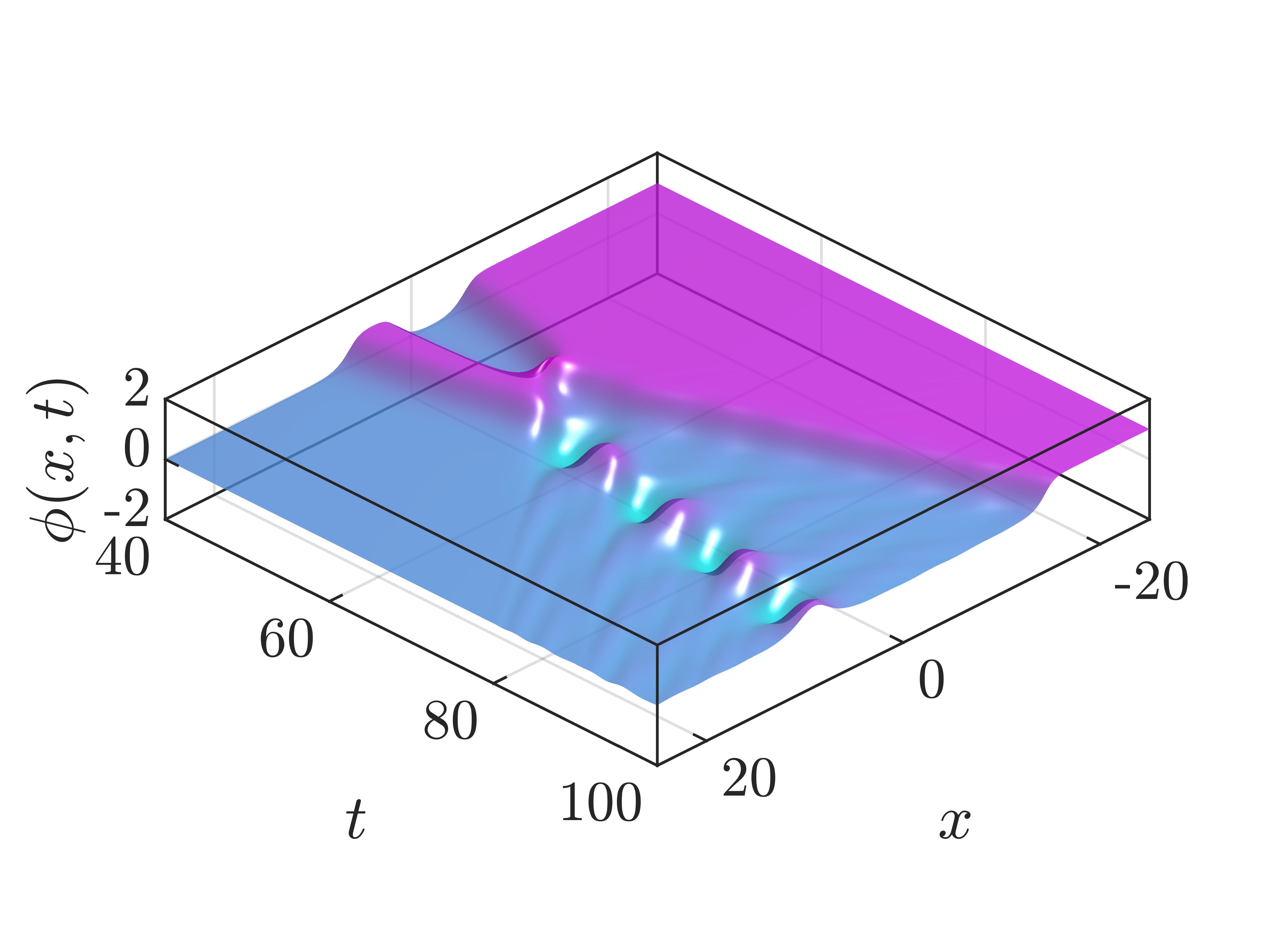}
\caption{$(1,0,1,0)$}
\label{fig_1010}
\end{subfigure}
\hfill
\begin{subfigure}[b]{0.325\textwidth}
\centering
\includegraphics[width=\textwidth]{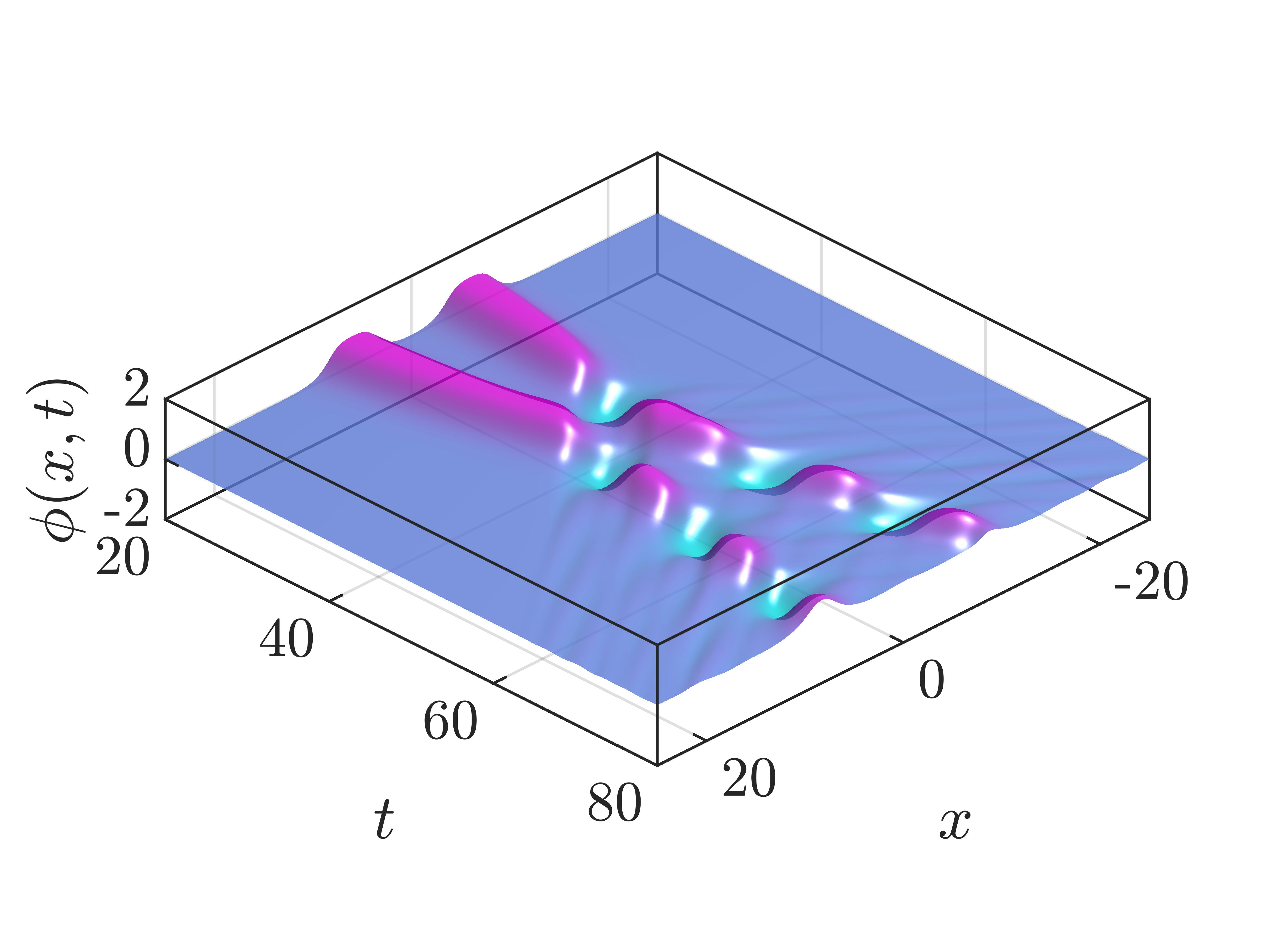}
\caption{$(0,1,0,1,0)$}
\label{fig_01010}
\end{subfigure}
\caption{The spacetime configuration for three- and four-kink collisions. The captions indicate the topology of the multi-kinks. The first row and the second row are fields for $\alpha=0$ and $\alpha=0.1$, respectively.}
\label{fig_multi}
\end{figure*}

The authors in Ref.~\cite{MoradiMarjaneh2017} have studied the multi-kink collision of the ordinary $\phi^6$ model in great detail. And it would be of interest to study how the results depend on the Lorentz violation. We follow the initial value settings in Ref.\cite{MoradiMarjaneh2017} to guarantee the collisions occur at the same point when $\alpha=0$ (the upper row in Fig.~\ref{fig_multi}). The parameter $\alpha$ is then slowly increased to study the impact of small Lorentz violation (the lower row in Fig.~\ref{fig_multi} shows the results of $\alpha=0$).


The three-kink collisions of topology $K\bar{K}K$ are presented in Fig.~\ref{fig_0101} and Fig.~\ref{fig_1010}. In the ordinary $\phi^6$ model, the field ends up with a right-moving kink and a small-amplitude left-moving bion. When $\alpha$ is turned on, the velocity of the kink and bion is decreased, whilst the amplitude of the bion is increased. In this case, part of the kinetic energy is transferred to the potential energy of the bion. The $\bar{K}K\bar{K}$ collisions are similar to those of $K\bar{K}K$, in the sense of forming bions. The four-kink collisions of $K\bar{K}K\bar{K}$ are shown in Fig.~\ref{fig_01010}. When $\alpha=0$, the four kinks collide and end up with two kinks and radiation in between. When $\alpha=0.1$, the collision creates two bions, and the bions travel slower than the kinks at $\alpha=0$. This tendency of forming bions is valid for small Lorentz violation parameters ($\alpha<0.1$). In this range, the kinks give more energy to the bions at higher $\alpha$.

It is now clear that a small $\alpha$ could lead to a very different final state, especially for the multi-kink collisions. In the example presented in Fig.~\ref{fig_multi}, a small Lorentz violation tends to suppress the speed of kinks and bions and uses the energy to build bions or make them stronger.

\section{Conclusions}
\label{sec_con}

In this study, we were concerned mainly with the kink scattering of a Lorentz-violating $\phi^6$ model. The static and boosted kink solutions were obtained in the spirit of Ref.~\cite{Barreto2006}. It was found that with Lorentz invariance broken, traveling kinks and antikinks became asymmetrical. The following analysis proved the stability of kink solutions. The eigenfunctions of the $\bar{K}K$ linear perturbation for $\alpha=0, 1, 2$, $0 \leq n \leq 4$ and separation $x_0=10$ were calculated, as shown in~\ref{fig_waveFunction}.

After the general remarks, the asymmetrical kink collisions are studied for both the $K\bar{K}$ and $(\bar{K}K)$ configurations. The Lorentz-violating $\phi^6$ model has no BW for $K\bar{K}$ collisions and no 3BW or higher bounce windows for $\bar{K}K$ collisions. The structure in Fig.~\ref{fig_structure} shows how $v_c$ decreases due to the presence of $\alpha$. We observed rising windows in the structure of $K\bar{K}$ scatterings, and the damping of the field value at the origin, or equivalently, the deviation of bions from the origin under Lorentz violation. To the author's knowledge, these two effects have not been reported yet.

We also discovered an empirical formula Eq.~\ref{eq_fit} of $v_c$ dependencies on $\alpha$ for the two topologies in the $\phi^6$ model and that of the $\phi^4$ model. One can study higher-order Lorentz-violating models to generalize the formula. We then studied the impacts of $\alpha$ on maximal energy densities. Due to the lacking of different modes of single kinks, the dependencies have no intricate structure as that of the $\phi^4$ model.

The impact of Lorentz violation on the multi-kink collisions was also studied. We found that the presence of small $\alpha$ would change the collision process. In those examples, the kinetic energy is transferred to forming bions or enlarging their amplitude.

\section{Acknowledgements}

The author is grateful to Yuan Zhong for helpful discussions.

\bibliographystyle{eplbib}
\bibliography{KKLV_phi6}

\end{document}